\documentclass[pdftex,12pt,a4paper]{article}

\usepackage[pdftex]{graphicx}
\usepackage[squaren]{SIunits}
\usepackage{amssymb}
\usepackage{fancyhdr}
\usepackage{enumerate}
\usepackage{amsmath}
\usepackage{color}
\usepackage{setspace}
\usepackage [latin1]{inputenc}
\usepackage[pdftex,bookmarks=true]{hyperref}
\usepackage{multirow}
\usepackage{authblk}
\usepackage{footmisc}

\usepackage{url}

\newcommand{\gfrac}[2]{\displaystyle\frac{#1}{#2}}

\newcommand{\dd}{\mbox{d}}

\newcommand{\LAT}{{\it Fermi}-LAT}

\newcommand{\HELAS}{{\it HELAS}}
\newcommand{\BH}{{\it Bethe-Heitler}}
\newcommand{\Jost}{{\it Jost}}
\newcommand{\emstandard}{{\it G4:emstandard}}

\newcommand{\livermorepola}{{\it G4:livermorepola}}
\newcommand{\penelope}{{\it G4:penelope}}
\newcommand{\EGSa}{{\it EGS0}}

\newcommand{\EGSc}{{\it EGS2}}

\begin{document}

\title{
$\gamma$-ray telescopes using conversions to $e^+e^-$ pairs: 
event generators, angular resolution and polarimetry
}
 
\author[1]{P.~Gros\thanks{philippe.gros at llr.in2p3.fr}}
\author[1]{D.~Bernard}
\affil[1]{LLR, Ecole Polytechnique, CNRS/IN2P3, 91128 Palaiseau, France}

\maketitle

\begin{abstract}
We benchmark various available event generators in {\it Geant4} and {\it EGS5} in the light of ongoing projects for high angular-resolution pair-conversion telescopes at low energy.
We compare the distributions of key kinematic variables extracted from the geometry of the three final state particles.
We validate and use as reference an exact generator using the full 5D differential cross-section of the conversion process.
We focus in particular on the effect of the unmeasured recoiling nucleus on the angular resolution.
We show that for high resolution trackers, the choice of the generator affects the estimated resolution of the telescope.
We also show that the current available generator are unable to describe accurately a linearly polarised photon source.
\end{abstract}



\section{Introduction}
\label{sec:intro}

$\gamma$-ray astronomy provides insight to understanding the non thermal processes in sources that undergo the most violent phenomena in the Universe, such as active galactic nuclei (AGN), gamma-ray bursts (GRB) and pulsars.
Compton telescopes are mainly sensitive below photon energies of a few MeV,
while existing pair telescopes are mainly sensitive above 100 MeV.
In between lies an energy range in which no high-sensitivity
measurements are available.

On the ``pair side'', the main issue is the strong degradation of the
angular resolution at low energy, which makes the rejection of the background noise from true
photon conversions  in the detector less efficient.
Several technologies are being considered to improve the angular resolution 
such as silicon wafer stacks (i.e., without tungsten converters)~\cite{TIGRE:2001,MEGA:2005,CAPSiTT:2010,Morselli:2014fua,Moiseev:2015lva,PANGU:2015,E-Astrogam:2016}, liquid noble-gas time projection chambers TPC
(argon \cite{Caliandro:2013kba}, or xenon \cite{Aprile:2008ft}) and emulsions \cite{Takahashi:2015jza}.

The two dominant contributions to the photon angular resolution are,
on the one hand the single track angular resolution of the two leptons, 
and on the other hand the impossibility to measure the momentum of the recoiling ion.
As the tracking performance improves, the missing recoil momentum becomes dominant.
It is therefore necessary to describe it accurately in detector simulations.

Additionally, the conversion azimuthal angle of the electron and the positron with respect to the flight direction of the incident photon can in principle enable the measurement of the polarisation fraction and the polarisation angle of the source~\cite{ber}.
With the low-density detectors presently under development, the track directions can be measured with enough precision before multiple scattering ruins the azimuthal information carried by the pair~\cite{Bernard:2013jea}.
The preliminary results of a recent experimental campaign on a particle accelerator beam are extremely encouraging in that regard~\cite{Gros:SPIE:2016}.

As the technology improves, we see the necessity to validate the event generators used in detector simulations.
In the present paper we reconsider the angular resolution due to the not measured recoil momentum with the new exact 5D event generator described in~\cite{Bernard:2012uf},
without any approximation in the calculation of the angular resolution,
extending the study to the lowest energies, very close to threshold,
and providing 68\%, 95\% and 99.7\% quantiles of the angular resolution.
We then characterise the behaviour of other available event generators in {\it Geant4} and {\it EGS5}, through a few key kinematic variables of the pair conversion process.
Finally, we examine the accuracy of these generators to describe the angular resolution and polarimetry potential for modern detectors with very fine tracking resolution.

\section{Kinematics}
\label{sec:kine}

We use the indices $i=+,-,r,\gamma$ for the quantities related respectively to the positron, electron, recoil nucleus and photon.
\iffalse
\begin{figure}[!h]
 \begin{center}
 \setlength{\unitlength}{0.47\textwidth}
 \begin{picture}(1,0.7)(0,0)
 \put(0,0){
\includegraphics[width=0.47\textwidth]{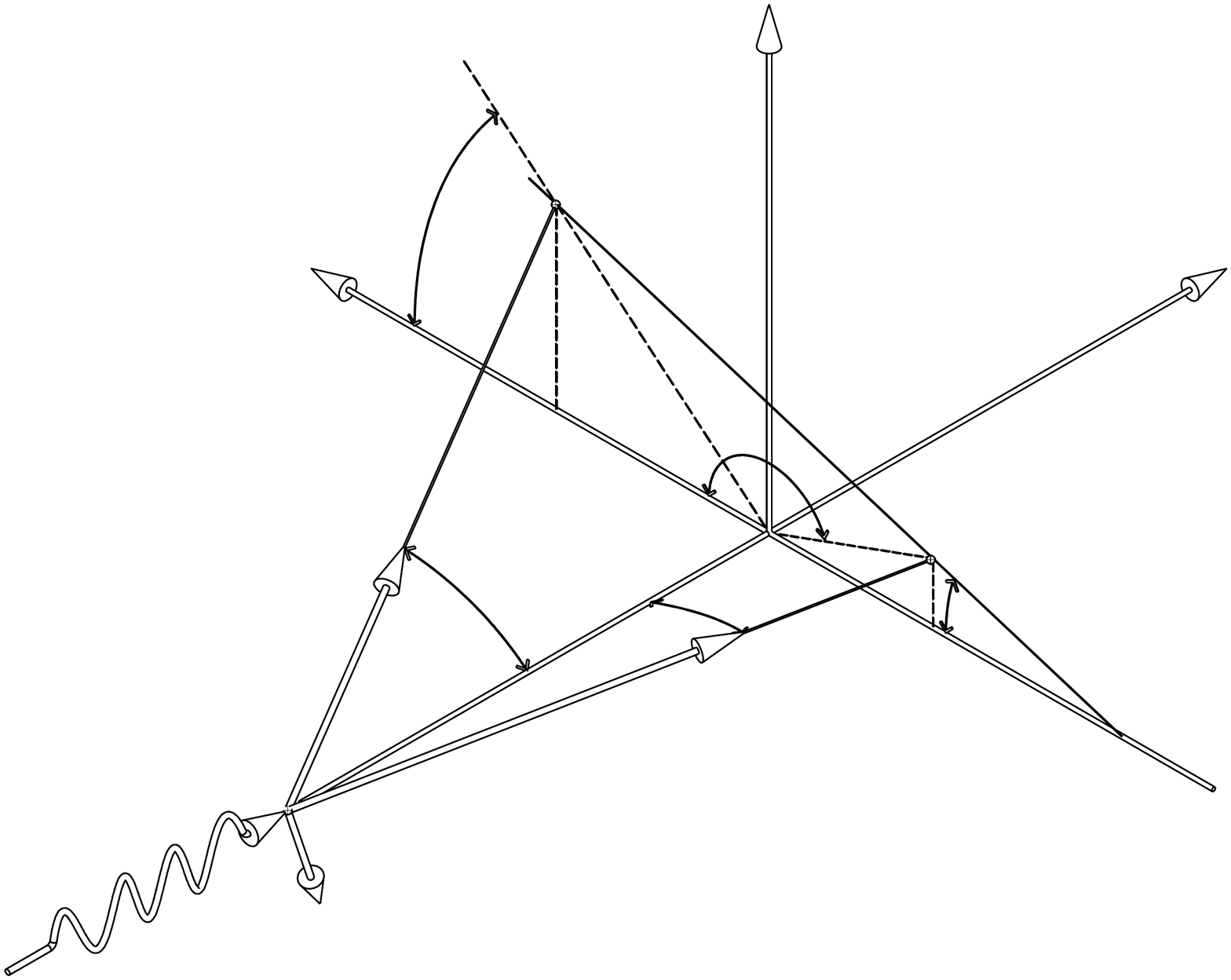}
}
 \put(0.27,0.5){$x$}
 \put(0.6,0.71){$y$}
 \put(0.97,0.5){$z$}
 \put(0.32,0.57){$\varphi_+$}
 \put(0.575,0.4){$\varphi_-$}
 \put(0.775,0.27){$\omega$}
 \put(0.41,0.29){$\theta_+$}
 \put(0.58,0.265){$\theta_-$}
 \put(0.3,0.3){$\vec{p_+}$}
 \put(0.59,0.2){$\vec{p_-}$}
 \put(0.3,0.02){$\vec{p_r}$}
 \put(0.03,-0.01){$\vec{k}$}
 \end{picture}
 \caption{Schema of a photon conversion.
 \label{fig:schema:angles}}
 \end{center}
\end{figure}
\else
\begin{figure}[!h]
 \begin{center}
 \setlength{\unitlength}{0.47\textwidth}
 \begin{picture}(1,0.7)(0,0)
 \put(0,0){
\includegraphics[width=0.47\textwidth]{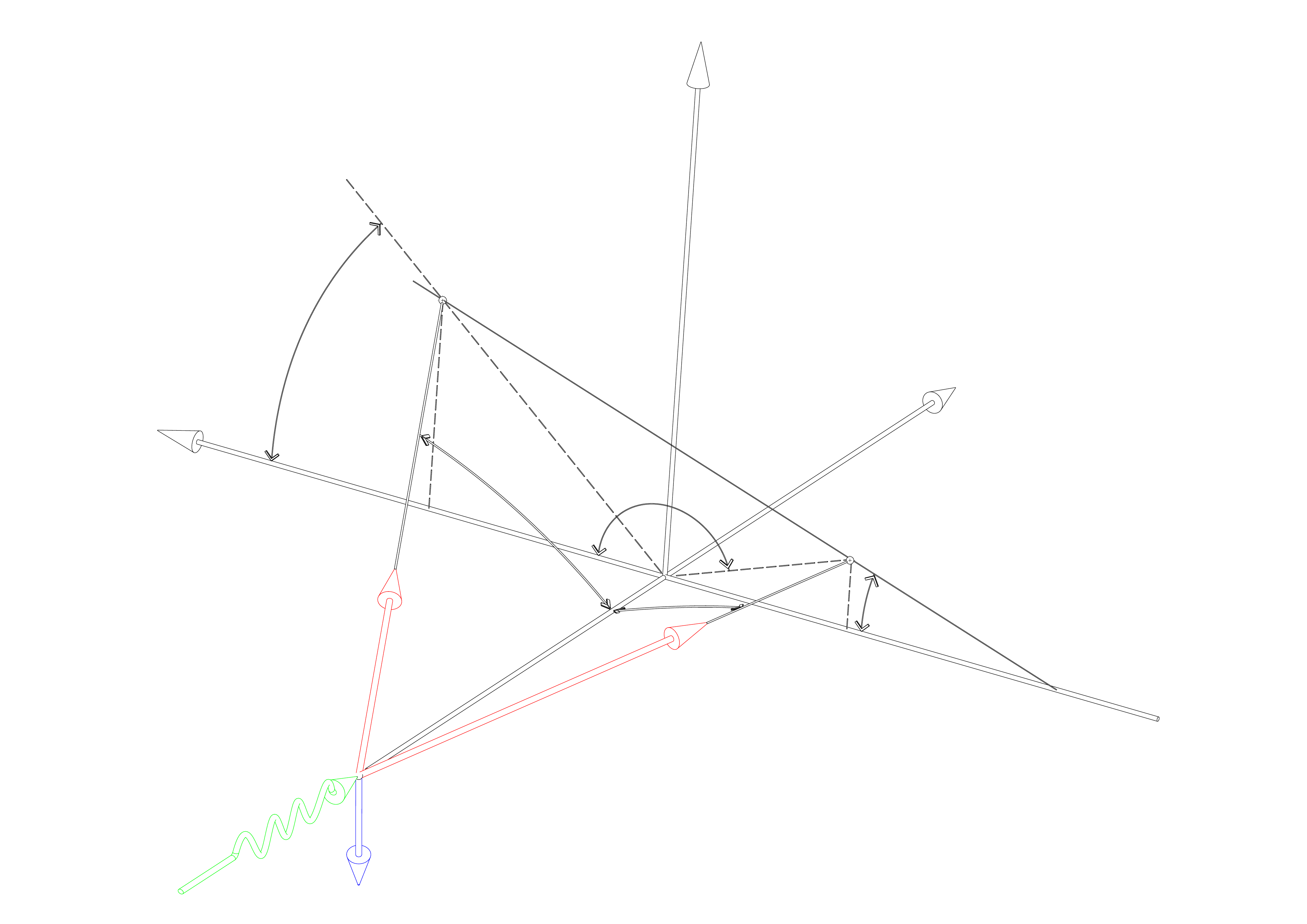}
}
 \put(0.13,0.4){$x$}
 \put(0.58,0.67){$y$}
 \put(0.7,0.44){$z$}
 \put(0.178,0.47){$\varphi_+$}
 \put(0.462,0.345){$\varphi_-$}
 \put(0.72,0.27){$\omega$}
 \put(0.39,0.24){$\theta_+$}
 \put(0.56,0.19){$\theta_-$}
 \put(0.22,0.2){\color{red}{$\vec{p_+}$}}
 \put(0.45,0.12){\color{red}{$\vec{p_-}$}}
 \put(0.32,0.02){\color{blue}{$\vec{p_r}$}}
 \put(0.08,-0.01){\color{green}{$\vec{k}$}}
 \end{picture}
 \caption{Schema of a photon conversion.
 \label{fig:schema:angles}}
 \end{center}
\end{figure}
\fi

\begin{itemize} 
\item $E_i,\vec{p_i}$ are the energy and momentum of the particle.
\item $\vec{u}_{i}$ is the propagation direction of the particle. By convention, we use $\vec{u_{\gamma}}$ as the $z$~direction.
\item $\varphi_{i}=\arctan\left(u_{i}^x/u_{i}^y\right)$ is the azimuthal angle of the particle.
\item $x_{i}=E_{i}/E_{\gamma}$ is the fraction of photon energy carried away by the particle. 
\item $q=|\vec{p_r}|$ is the momentum transferred to the recoil nucleus.
\item $\theta_{+-}$ is the pair opening angle, that is, the angle between the electron and positron momenta at conversion point, $$\theta_{+-} = \arccos(\vec{u}_{+} \cdot \vec{u}_{-}).$$
\item $\theta$ is the angle between the incident photon and the observed direction $\vec{u}_{\rm{pair}}$ obtained from the summed 4-vectors of the electron and positron. $\theta = \arccos{u_{\rm{pair}}^z}$.
We discuss in Sect.~\ref{subsec:ang:res} the estimation of $\theta$ when the full 4-vector information is not available.
\item $\phi$, the conversion azimuthal angle, is defined as~\cite{Gros:2016:azimuthal} $$\phi = \frac{\varphi_{+}+\varphi_{-}}{2}.$$
\end{itemize} 

\section{Validation of the simulation of the full 5D differential cross-section}
\label{sec:validation}
  
With the prospect of having a pair telescope sensitive to photons of much lower energies than at present, there is a need for a conversion-to-pair event generator that is:
\begin{itemize} 
\item 
exact down to threshold, that is, without any low-energy nor small-angle approximations, 
\item 
yielding a sampling of the true, five-dimensional (5D) differential cross section, that is, not a simple product of 1D projections, 
\item 
allowing the generation of conversion events by a linearly polarised beam.
\end{itemize} 

We developed a generator~\cite{Bernard:2013jea}, based on the
BASES/SPRING event generator~\cite{Kawabata:1995th}, that instantiate
the VEGAS Monte Carlo method \cite{Lepage:1977sw}: the differential
cross section is tabulated in a 5-dimensional space on a grid that has
been optimized to minimize its uncertainty.
Events are then taken at random from the exact differential cross
section using the acceptance-rejection method from the tabulated
mock-up.
Two methods were used to compute the 5D differential cross section.
The Bethe-Heitler analytic expression~\cite{Heitler1954} (that includes only the two dominant diagrams, an approximation that is valid for nuclear conversion and for high-energy triplet conversion), we then refer to it as the \BH\ model;
A full diagram computation using the HELAS amplitude calculator~\cite{Murayama:1992gi}, to which we refer as the \HELAS\ model.
The polarised form of the Bethe-Heitler 5D differential cross section is from Ref.~\cite{May1951}, after a term was corrected by a factor of 2~\cite{jau};
the expressions that are used can be found in Ref.~\cite{Gros:2016:azimuthal}.

We confronted the calculations of that new generator to the analytic results on 1D projections published in the past and used it to study the performance of an actual TPC telescope and polarimeter~\cite{Bernard:2013jea}.

The generator we developed relies on a relatively complex combination of packages.
Our first step is to cross-validate the simulation by comparing three different approaches:
\begin{itemize}
\item A full simulation using Feynman diagram amplitudes calculated with HELAS, and 5D generator using  BASES/SPRING.
\item A simulation using the Bethe-Heitler approximation (very accurate in case the target is a nucleus much heavier than the electron), and 5D generator using   BASES/SPRING.
 Bethe and Heitler neglected the diagrams for which the incident photon has a vertex with the ion (diagrams (c) and (d) of Fig. 1 of Ref. \cite{Bernard:2013jea}).
\item An analytic distribution of the recoil momentum of the nucleus obtained by Jost {\em et al.}~\cite{Jost:1950zz} by integrating the BH differenttila cross section over other variables.
\end{itemize}

\begin{figure}[\plotflag]
\begin{center}
  \includegraphics[width=0.45\linewidth]{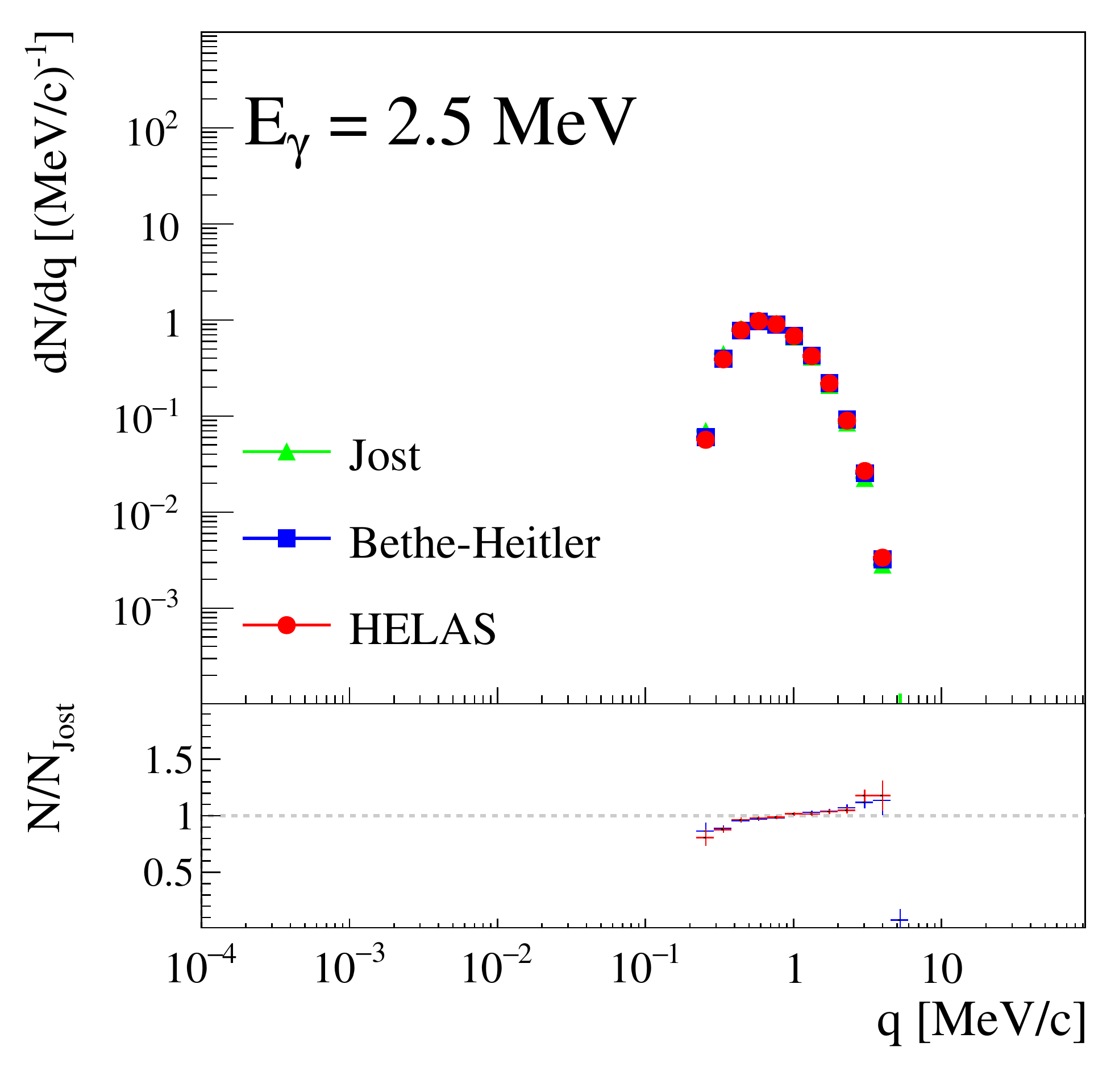}
  \includegraphics[width=0.45\linewidth]{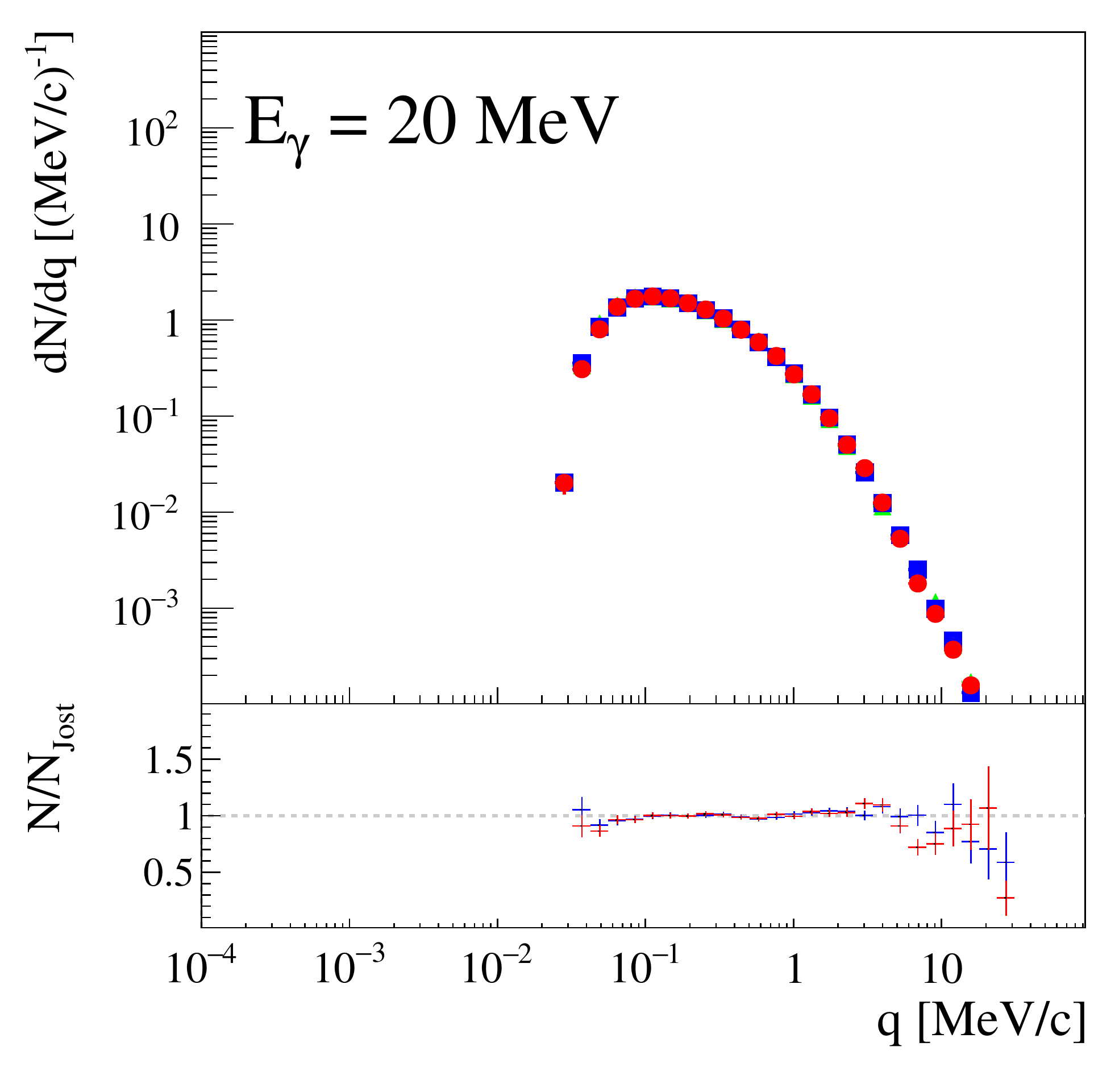}
  \includegraphics[width=0.45\linewidth]{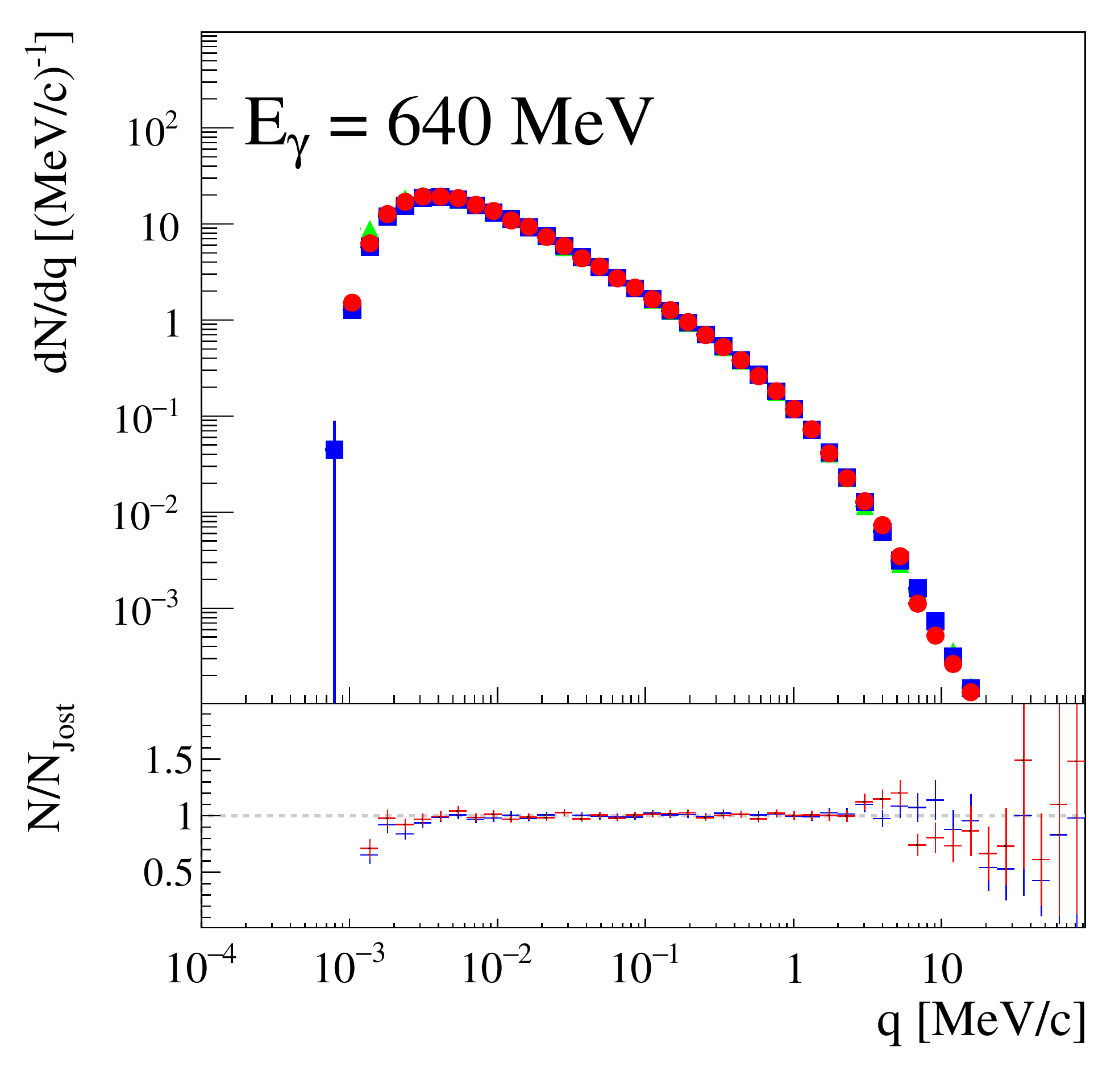}
  \caption{
    Examples of distribution of the momentum transfer $q$ for different energies (2.5\,MeV, 20\,MeV, 640\,MeV), without atomic screening form factor.
    The three models are fully consistent, as expected (as shown by the values normalised to Jost's analytic formula.
    \label{fig:valid:Q:examples}
  }
\end{center}
\end{figure}

We first compare the distributions of the recoil momentum $q$ transferred to the nucleus.
Since the recoil nucleus is cannot observed, $q^2$ is computed from the momenta $p_{\gamma}$, $p_{+}$ and $p_{-}$ of the incident photon and of the produced leptons as:
$ q^2 = \left| \vec{p_{\gamma}} - (\vec{p_{+}} + \vec{p_{-}}) \right|^2 $.
Figure~\ref{fig:valid:Q:examples} shows an example of the distributions for various photon energies.
The three distributions are perfectly consistent, which is a two-fold validation:
\begin{itemize}
\item a cross validation of the differential cross sections computed by HELAS and by the Bethe-Heitler expression;
\item a validation of our use of the BASES/SPRING  generator for gamma conversion.
\end{itemize}

\begin{figure}[\plotflag]
\begin{center}
  \includegraphics[width=0.42\linewidth]{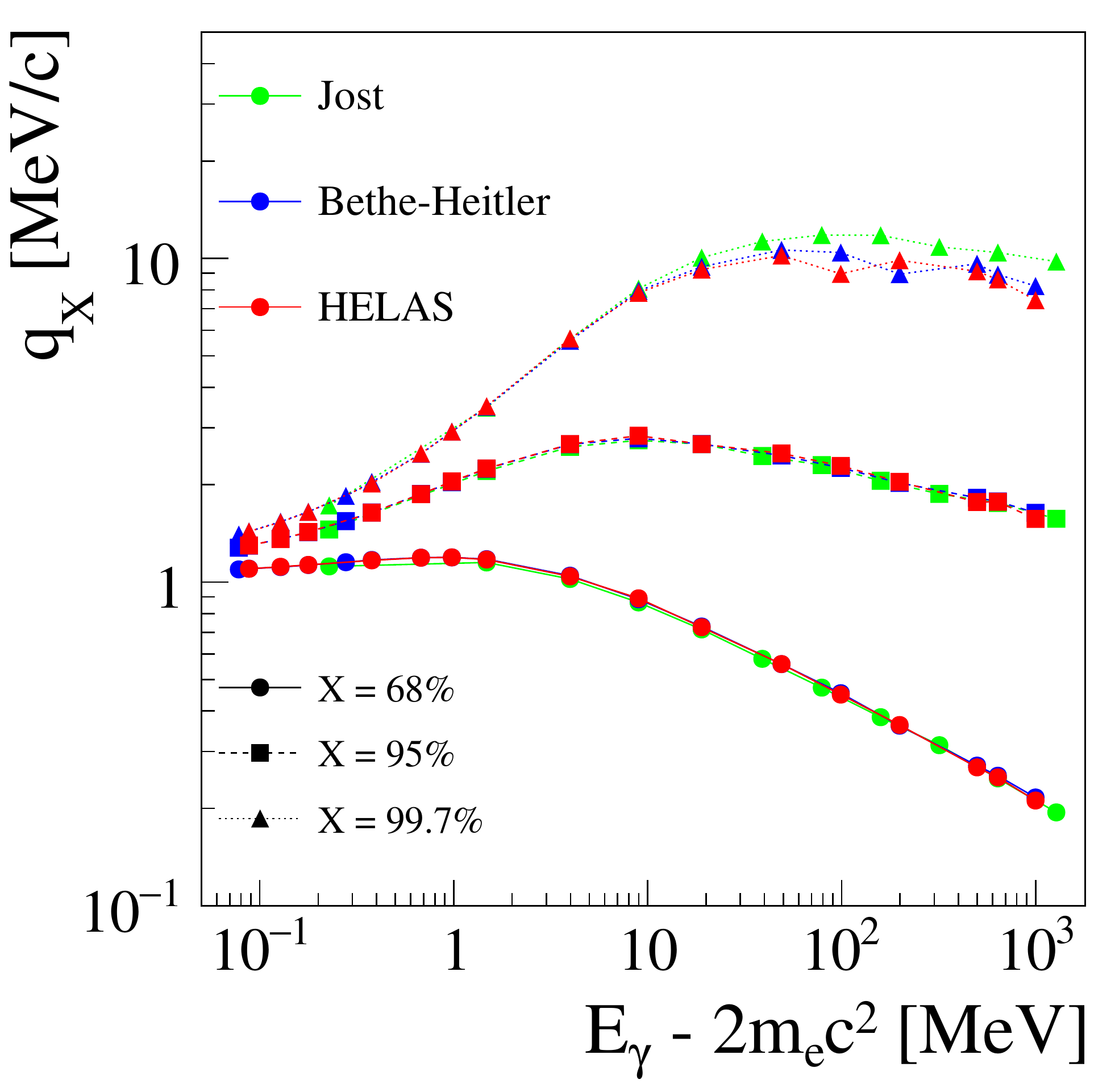}
  \includegraphics[width=0.42\linewidth]{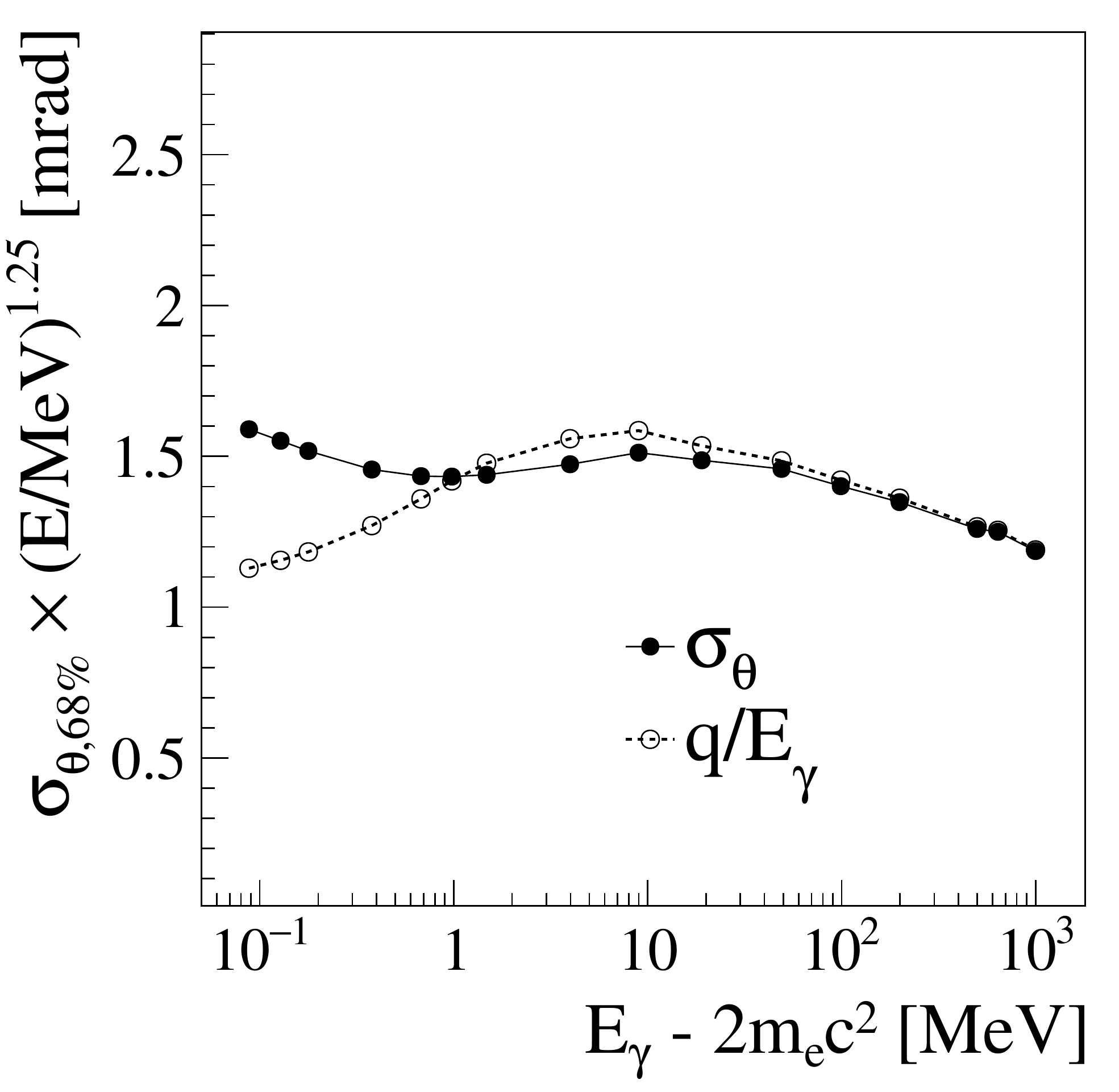}
  \caption{
    Left: containment value $q_{X}$ as a function of the photon energy, for different value of $X$ (dots: 68\,\%, squares: 95\,\%, triangles: 99.7\,\%).
    The different values of $X$ change the quantity $q_X$, but are qualitatively similar.
    The Bethe-Heitler, the HELAS and the Jost models
     give identical results, regardless of the containment value.
    Right: Comparison of the angular resolution $\sigma_{\theta,68\%}$ (full circles) with the high energy approximation $q_{68\%}/E_{\gamma}$ (open circles).
    The approximation is nearly exact above 40\,\mega\electronvolt, and within 10\,\% down to 1\,\mega\electronvolt above the conversion threshold.
    A scaling of $E_{\gamma}^{1.25}$ is used to make the comparison easier.
    \label{fig:valid:Q:modelCont}
  }
\end{center}
\end{figure}

As the $q$ distributions extend over several orders of magnitude, we obtain an estimate of their magnitude from the containment value $q_{X}$, defined as the momentum value such that a given fraction $X$ of the events have a recoil momentum smaller than $q_{X}$.
We use the containment values $X=$68\,\%, $X=$95\,\% and $X=$99.7\,\% that correspond approximately to 1$\sigma$, 2$\sigma$ and 3$\sigma$ for a Gaussian statistics.
Figure~\ref{fig:valid:Q:modelCont} shows the variation of $q_X$ with $E_{\gamma}$ for the three models.
Once again we see that the three models are in perfect agreement over the considered energy range ($< 1\,\giga\electronvolt$).

The recoil momentum $q$ is crucial due to its close relation to the photon angular resolution $\sigma_{\theta}$ of the detector when the recoil nucleus cannot be measured as is the case for nuclear conversion.
In that case, the direction is estimated from the electron-positron pair (sum of 4-vectors).
At high energy, the angle between the photon and the pair can be approximated as $\theta \approx q/E_{\gamma}$~\cite{Bernard:2012uf}.
Figure~\ref{fig:valid:Q:modelCont} right, shows that this approximation is exact above 40\,\mega\electronvolt, and within 10\,\% down to 1\,\mega\electronvolt\ above the conversion threshold.
A scaling factor of $E_{\gamma}^{1.25}$ is used to make the comparison easier
\cite{Bernard:2012uf}.

\begin{figure}[\plotflag]
\begin{center}
  \includegraphics[width=0.42\linewidth]{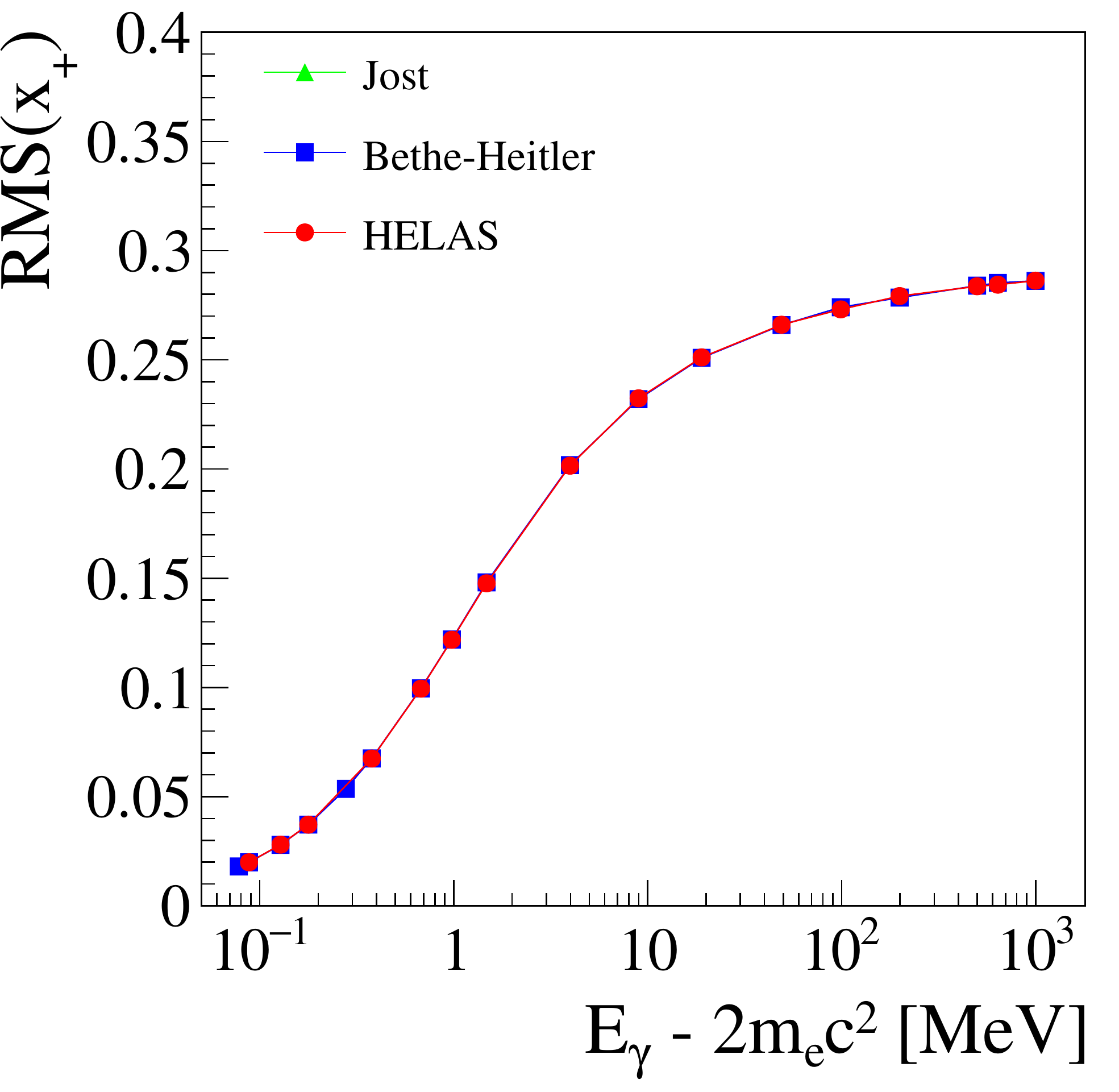}
  \includegraphics[width=0.42\linewidth]{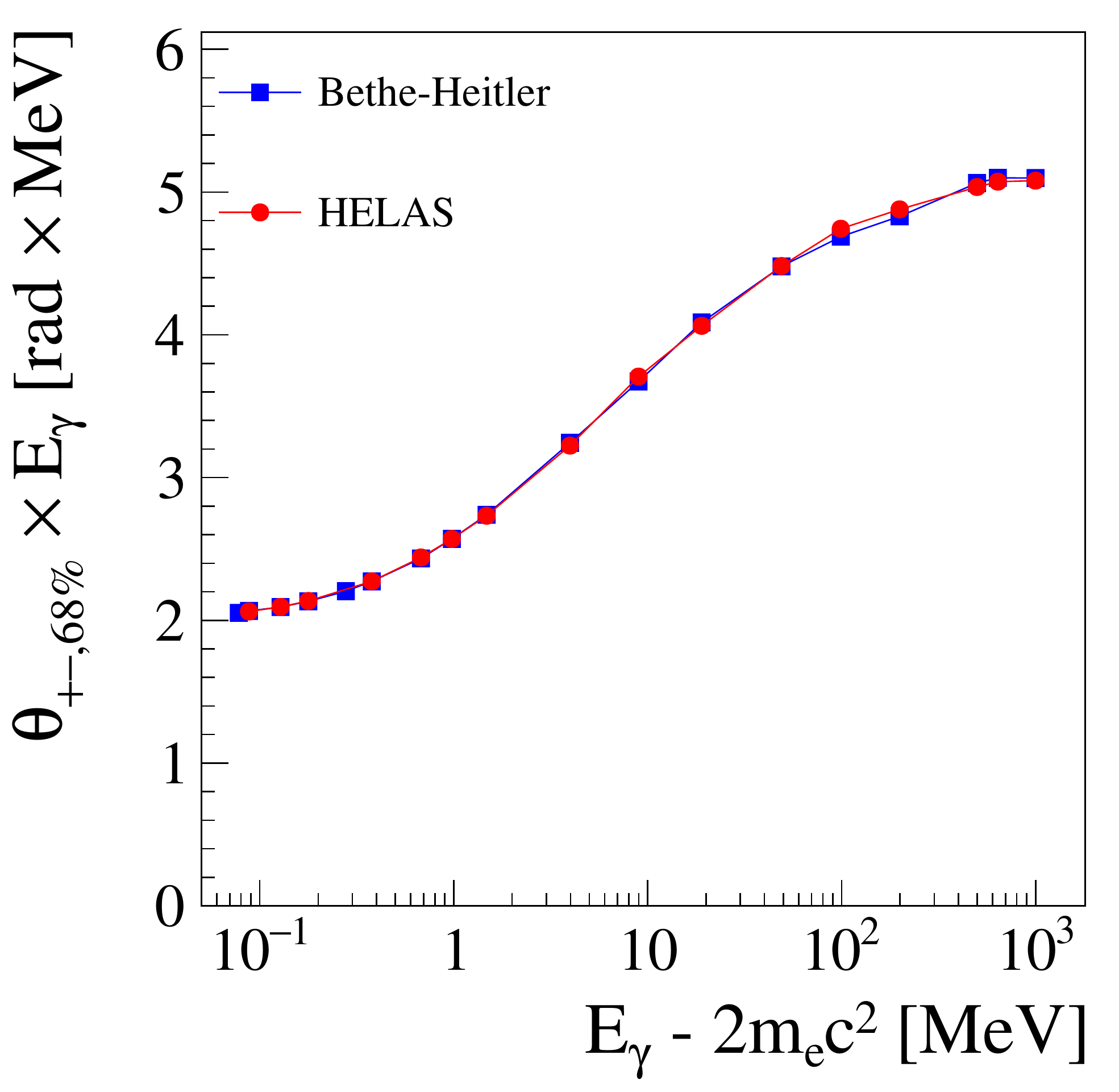}
  \caption{
    Comparison of the variables $x_+$ and $\theta_{+-}$ between \BH\ and \HELAS.
    The values are practically identical for both models.
    The full $x_+$ and $\theta_{+-}$ distributions confirm the agreement (plots not shown).
    \label{fig:valid:other:model}
  }
\end{center}
\end{figure}

Finally, we compare the distribution of the other kinematic variables $x_+$ and $\theta_{+-}$ for the \HELAS\ and \BH\ models.
The \Jost\ model only describes $q$.
We reduce the data by using the RMS of $x_+$, and the 68\,\% containment value of $\theta_{+-}$.
Figure~\ref{fig:valid:other:model} shows that over the full energy range, the two models are perfectly consistent.
An examination of the full $x_+$ and $\theta_{+-}$ distributions confirms the agreement (plots not shown).
At high energy, the distribution of $x_+$ is nearly flat between 0 and 1, which is reflected by the value of the RMS close to $1/\sqrt{12} \approx 0.29$.

We find that the approximation of the \BH\ model is in complete agreement with the full Feynman amplitude calculations in the \HELAS\ model.
This confirms that the simulation program we used is correct.
In the rest of this work, we only use the \HELAS\ model.

\begin{figure}[\plotflag]
\begin{center}
  \includegraphics[width=0.52\linewidth]{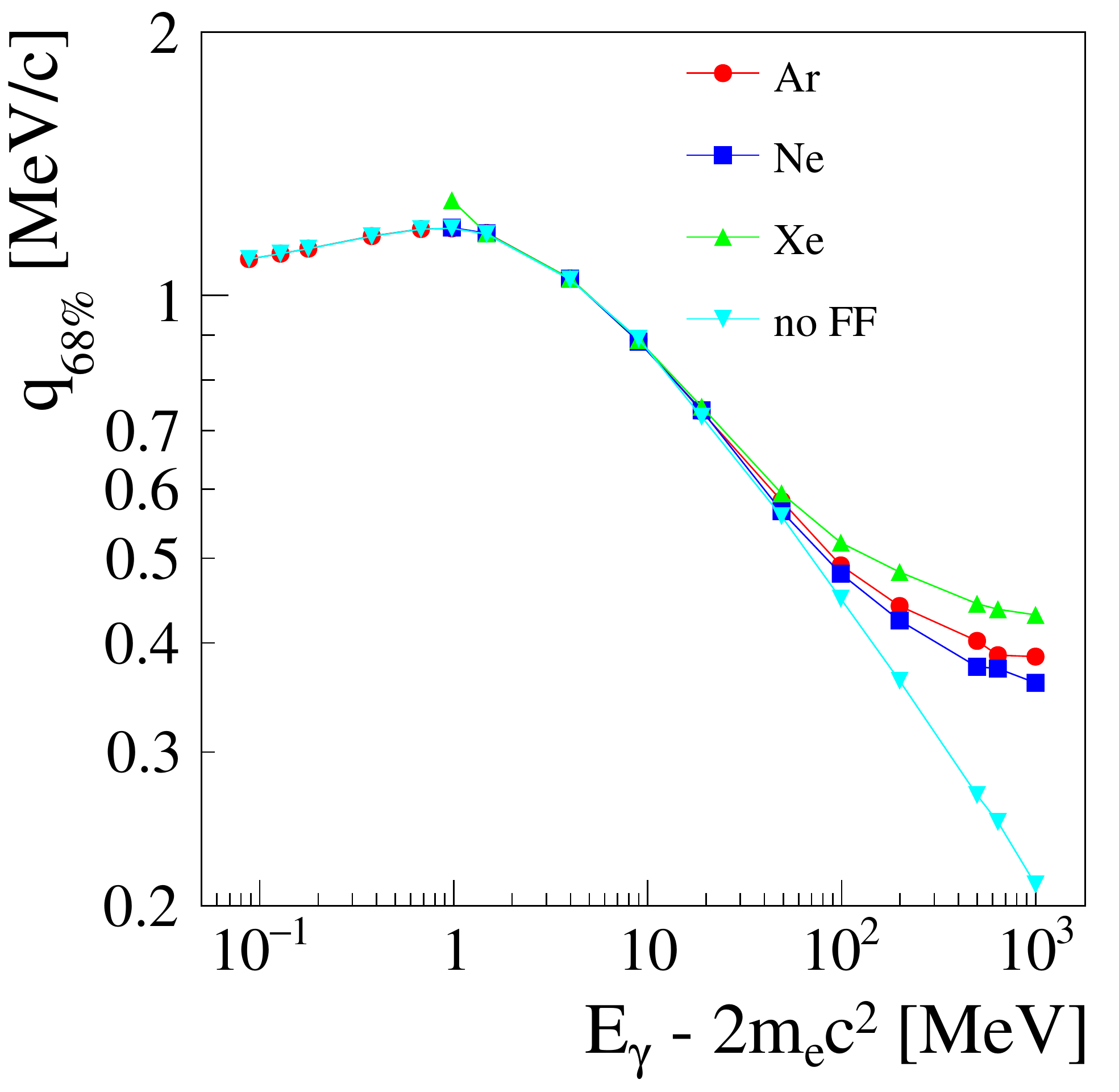}
  \caption{
    The recoil momentum containment value $q_{68\%}$ as a function of the photon energy, for different gas media ({\it Argon}, {\it Neon} or {\it Xenon}), and neglecting the form factor ({\it no FF}).
    The form factor affects the recoil momentum distribution significantly only for photon energies above 20\,MeV.
    \label{fig:valid:Q:gas}
  }
\end{center}
\end{figure}

We include the screening effects of the atomic structure by multiplying the differential cross-section by a form factor~\cite{Bernard:2013jea}.
Figure~\ref{fig:valid:Q:gas} shows the influence of the form factor on the recoil momentum $q$. 
It is negligible at low energies, but becomes strong above a few 10\,MeV.
This means that at high energies, the nature of the conversion medium should be taken into account for estimating the angular resolution.

\section{Comparison with other generators (Geant4 and EGS5)}
\label{sec:comparison}

\subsection{Event generators}
\label{subsec:evt:gen}

Geant4 is a widely used software for the simulation of detector physics~\cite{Geant4,Allison:2016lfl}.
We use version~10.02.01, enabling only the pair conversion process, and ignoring every other physics effect.
We considered the following pair conversion processes in the low energy electromagnetic physics lists:

\begin{description}
\item[emstandard] using the class \texttt{G4BetheHeitlerModel}. 
$E_{\pm}$ is sampled using the Bethe-Heitler cross section with Coulomb correction.
The recoil of the nucleus is neglected so that the electron-positron pair is in a plane that contains the photon direction (``coplanar generation'', i.e. $\varphi_+ - \varphi_- = \pm \pi$).
In that case, the photon direction should be perfectly known from the electron and positron 4-vectors.
However, some intrinsic uncertainty on the photon direction remains, due to the high-energy and small angle approximations used.
\item[livermore] using the class \texttt{G4LivermoreGammaConversionModel}.
The kinematics are identical to {\it emstandard}.
Only the total cross-section is different, which does not affect the present work.
This model is therefore equivalent to {\it emstandard}  and the results that we obtained with it are not presented.
\item[livermorepola] using the class \texttt{G4LivermorePolarizedGammaConversionModel}.
The asymptotic behaviour at high energy is based on the results in~\cite{Depaola:2000qd}.
The low energy behaviour is not documented.
\item[penelope] using the class \texttt{G4PenelopeGammaConversionModel}.
$E_{\pm}$ is sampled from~\cite{Baro} with high-energy Coulomb correction and atomic screening.
The direction of the electron and of the positron are taken at random independently\footnote{\label{note:violation}These models were developed and tuned to accurately describe electromagnetic showers, even though individual interactions violate the conservation of energy-momentum}.
\end{description} 

The EGS (Electron-Gamma Shower) code system is a general purpose package for the Monte Carlo simulation of the coupled transport of electrons and photons~\cite{EGS5,Hirayama:2005zm}.
We use EGS5 version~1.0.6. 
The conversion is coplanar to the photon direction, $\varphi_+ - \varphi_- = \pm \pi$.
The polar angles of the electron and of the positron are taken independently\footref{note:violation}.

The method used for determining the polar angles in EGS5~\cite{Hirayama:2005zm,Bielajew} is determined by a parameter IPRDST:
\begin{itemize}
\item IPRDST$=0$ (default) the two polar angles are fixed to the same value $\theta_{+} = \theta_{-} = m/E$.
\item IPRDST$=1$, the polar angle of the two leptons are taken, independently, from the leading order factor of the Sauter-Gluckstern-Hull expression (first Born approximation, unscreened-point-nucleus, negligible nuclear recoil), formula 3D-2000 of \cite{Motz:1969ti}, that is:
 $\gfrac{\dd p}{\dd \theta_\pm} = \gfrac{\sin{\theta_\pm}}{2 p_\pm (E_\pm - p_\pm \cos\theta_\pm)^2}$
\item IPRDST$=2$, the polar angle of the two leptons are taken, independently, from the Schiff expression (first Born approximation, extremely relativistic, negligible nuclear recoil, small polar angles, exponential screening) formula 3D-2003 of \cite{Motz:1969ti}.
 It should be noted that only the angular distribution part of the Schiff differential cross section is used.
\end{itemize}

We found similar results with IPRDST$=1$ and IPRDST$=2$, so that here we only show IPRDST$=0$ (\EGSa) and IPRDST$=2$~(\EGSc).

The generators used in this study are summarised in Table~\ref{tab:models}.

\begin{table}[h] 
\small
\begin{tabular}{ |l|l|l|l| } 
\hline
Name & Model & Generator & Ref \\
\hline
\hline
\HELAS\ & HELAS Feynman amplitudes & BASES/SPRING & \cite{Bernard:2013jea} \\
\hline
\BH & Bethe-Heitler & BASES/SPRING & \cite{Bernard:2013jea} \\
\hline
\hline
\emstandard & G4BetheHeitler & Geant4 10.02.01 & \cite{Urban,Tsai:1973py} \\ 
\hline
\livermorepola & G4LivermorePolarizedGammaConversion & Geant4 10.02.01 & \cite{Depaola:2000qd} \\ 
\hline
\penelope & G4PenelopeGammaConversion & Geant4 10.02.01 & \cite{Baro} \\
\hline
\hline
\EGSa & egs5, IPRDST$=0$ & egs5 1.0.6 & \cite{Hirayama:2005zm} \\
\hline
 \EGSc & egs5, IPRDST$=2$ & egs5 1.0.6 & \cite{Hirayama:2005zm,Motz:1969ti} \\
\hline
\end{tabular}
\caption{Summary of the models considered in this work. \label{tab:models}}
\end{table}

\subsection{Kinematic variables $q$, $x_+$ and $\theta_{+-}$}
\label{subsec:kin}

We use Argon as a conversion medium, but the form factor does not seem to be taken into account in the detail of the kinematics of the final state neither in Geant4 nor in EGS5.
The nature of the medium only affects the total cross-section.
We therefore use the \HELAS\ model without form factor {\it (no FF)} as a reference.

\begin{figure}[\plotflag]
\begin{center}
  \includegraphics[width=0.45\linewidth]{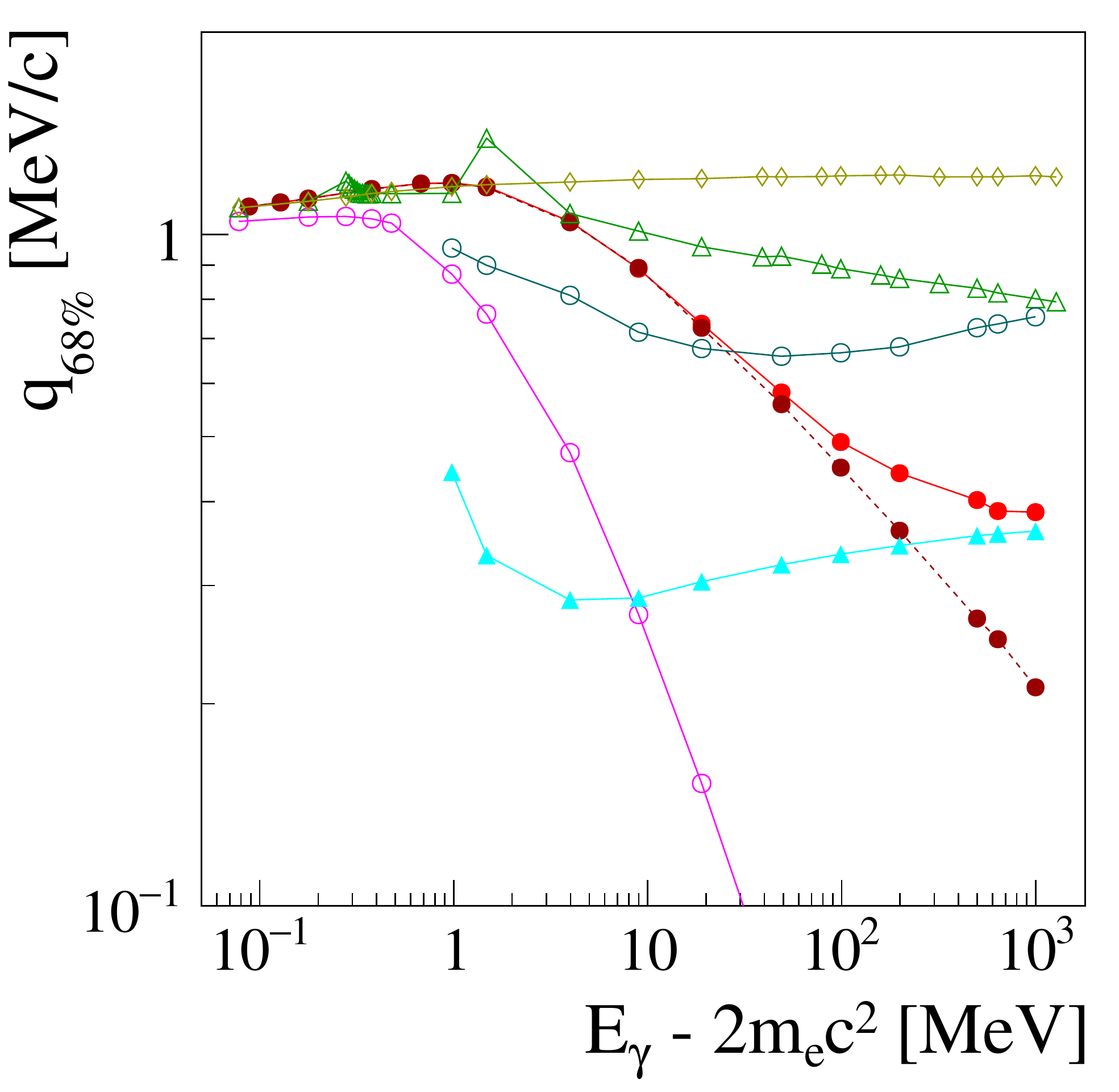}
  \includegraphics[width=0.45\linewidth]{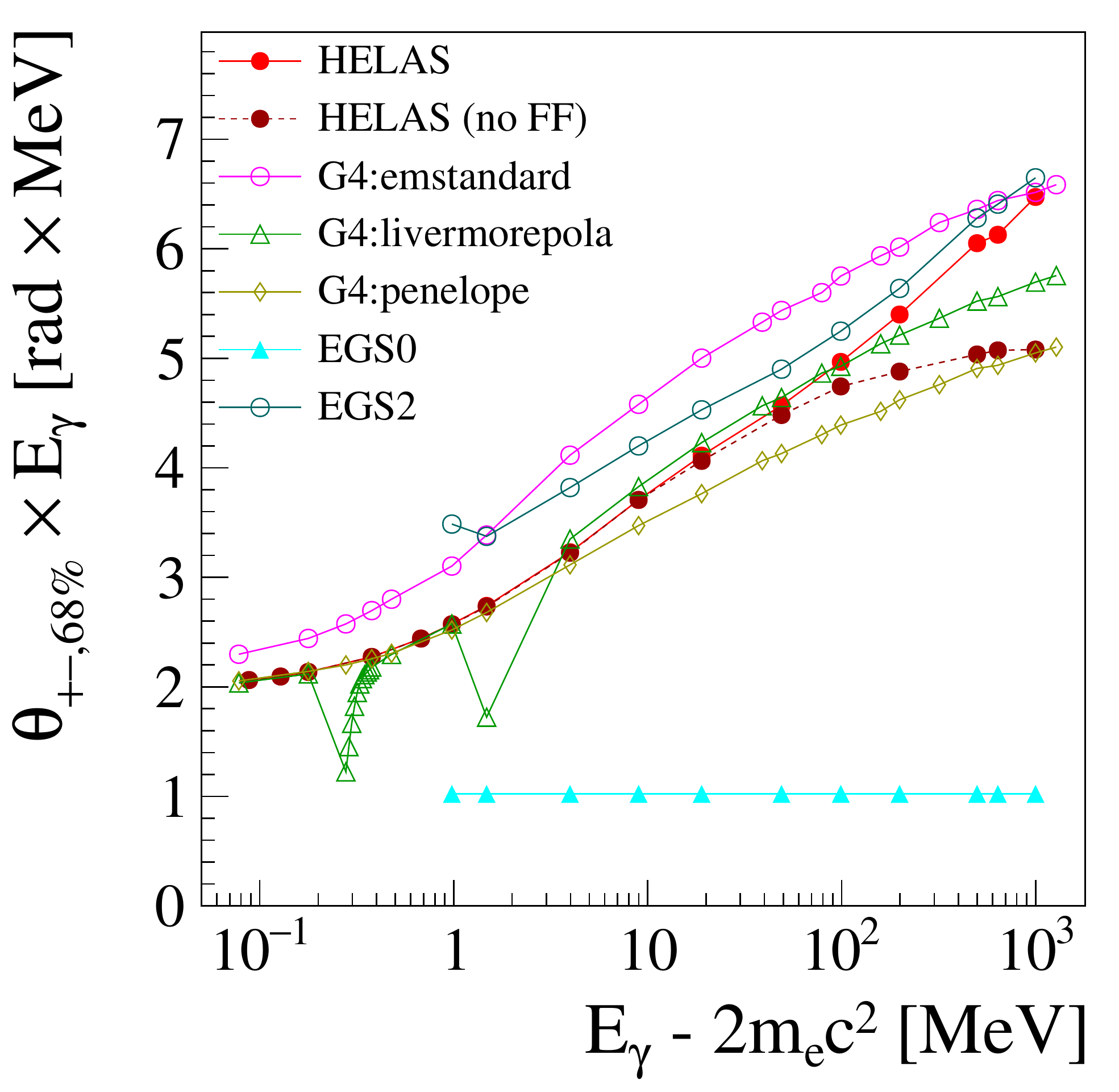}
  \includegraphics[width=0.45\linewidth]{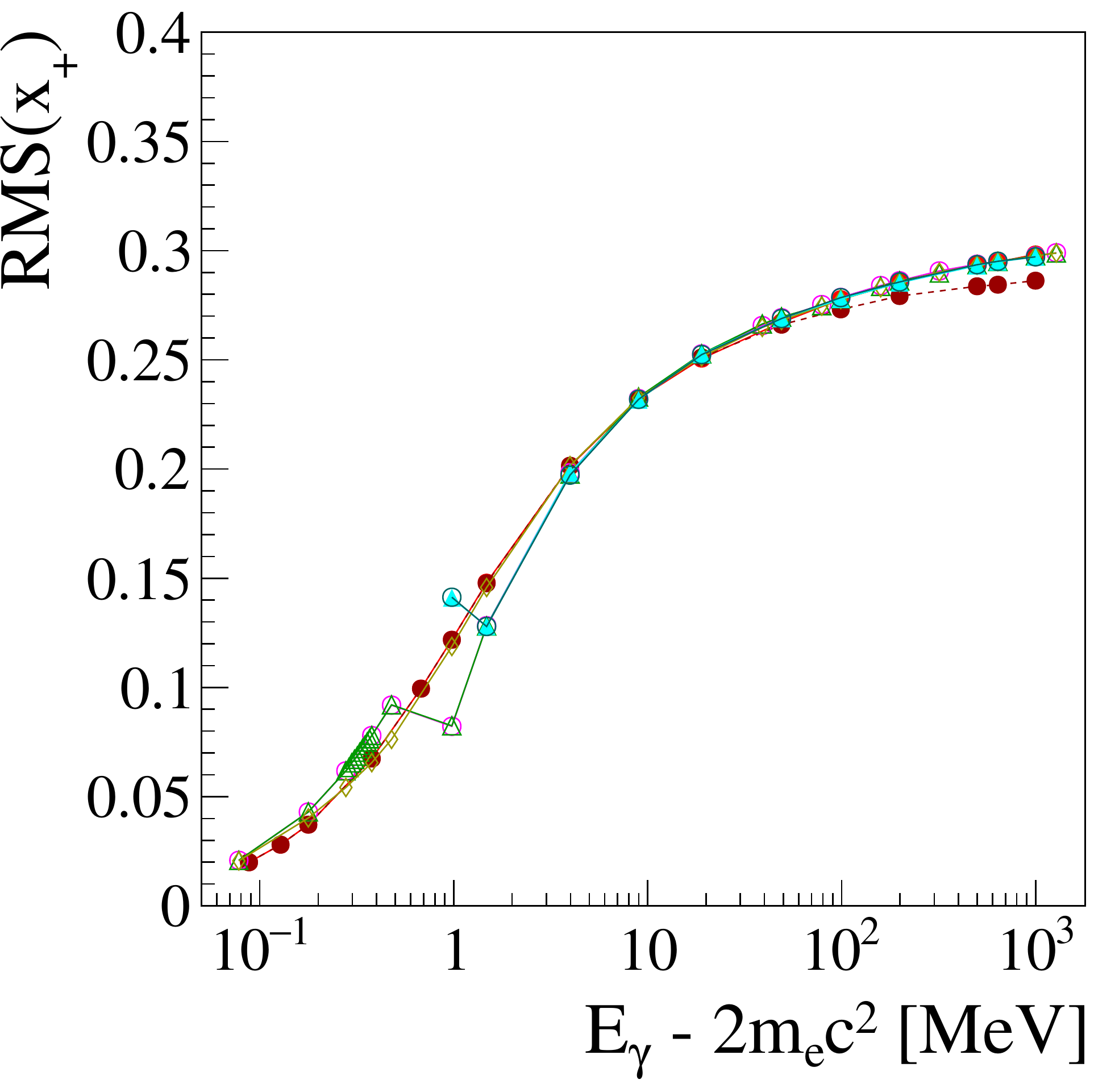}
  \caption{
    Comparison of the variables $q$, $\theta_{+-}$ and $x_+$ for the various models.
    The values of $x_+$ are consistent in all the models (with two exceptions at the border of regime changes around 1.3\,\mega\electronvolt\ and  2\,\mega\electronvolt\ for \livermorepola).
    The distributions of $q$ and $\theta_{+-}$ vary considerably between models.
    \label{fig:compG4:model}
  }
\end{center}
\end{figure}

Figure~\ref{fig:compG4:model} shows the variations of the different kinematic variables for each model.
We see that, as expected, \emstandard\ and \EGSa\ describe the recoil momentum $q$ poorly.
\livermorepola\ has two features around $E_{\gamma}=1.3\,\mega\electronvolt$ and $E_{\gamma}=2\,\mega\electronvolt$ which correspond to two regime changes.
Except for these features, the energy fraction $x_+$ is described consistently in all the models.
The distributions of $\theta_{+-}$ and especially of $q$, which are important for the performance of pair telescopes, are very different for all the models.

\subsection{Angular resolution}
\label{subsec:ang:res}

\begin{figure}[\plotflag]
\begin{center}
  \includegraphics[width=0.45\linewidth]{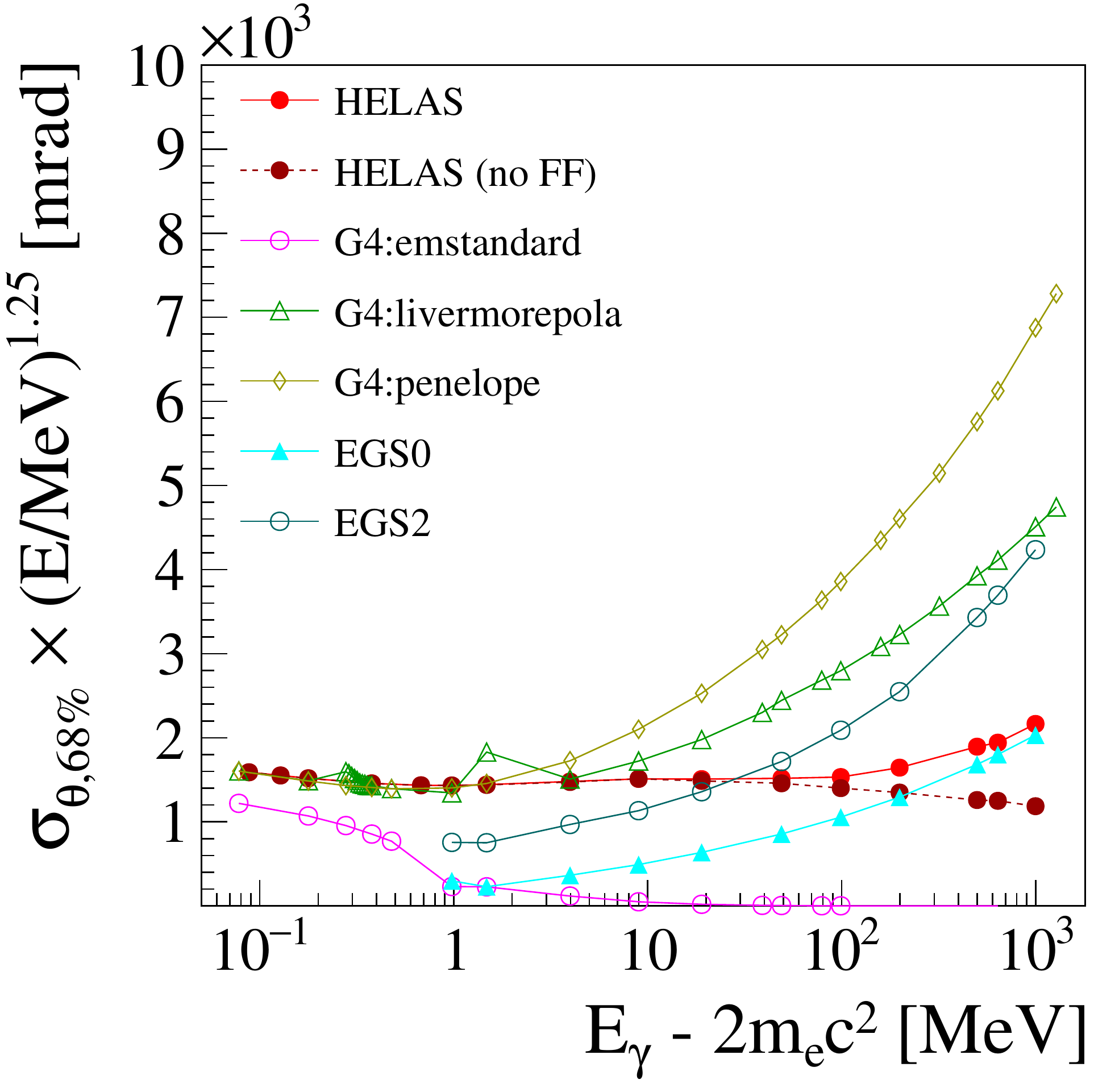}
  \includegraphics[width=0.45\linewidth]{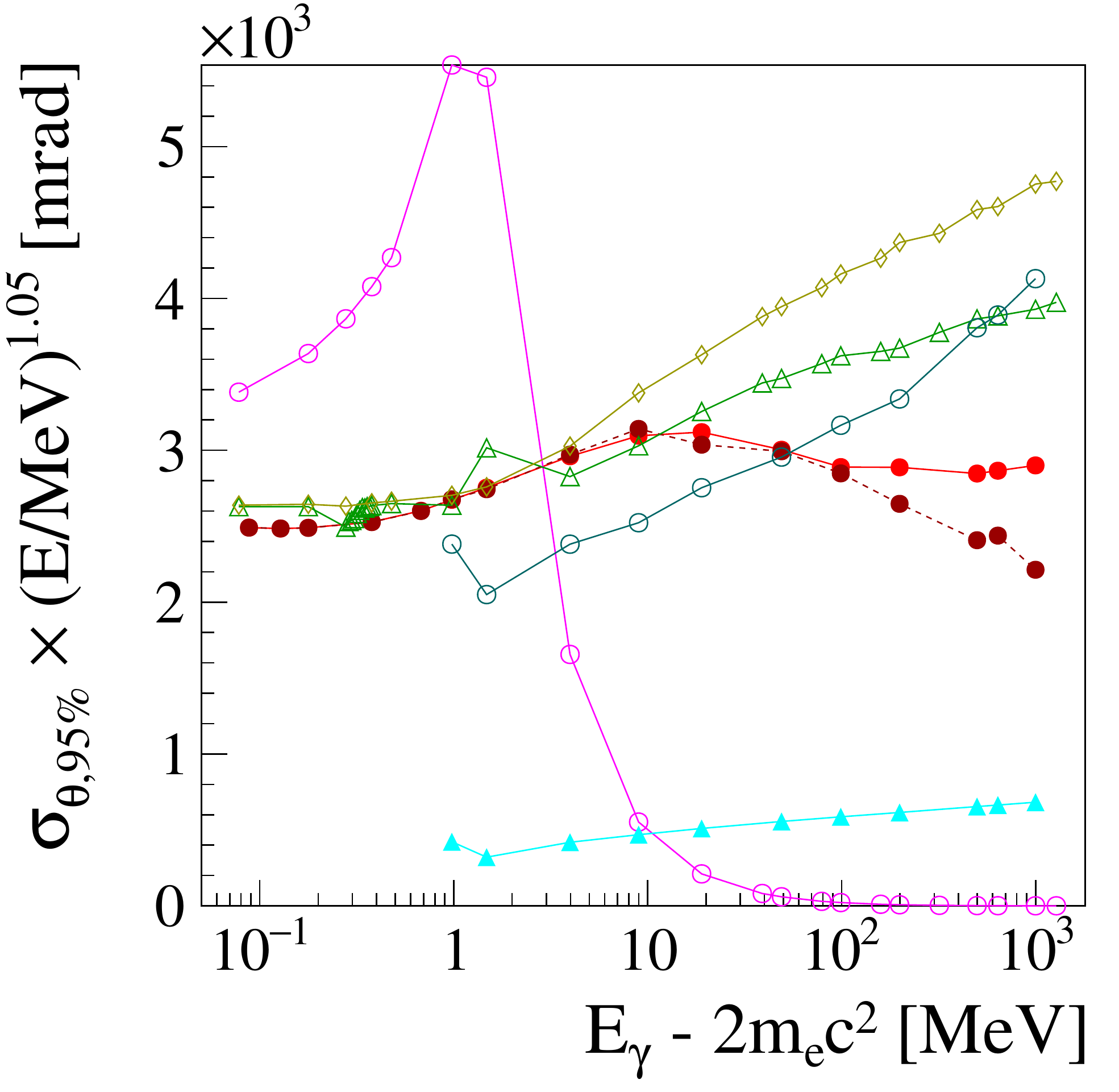}
  \includegraphics[width=0.45\linewidth]{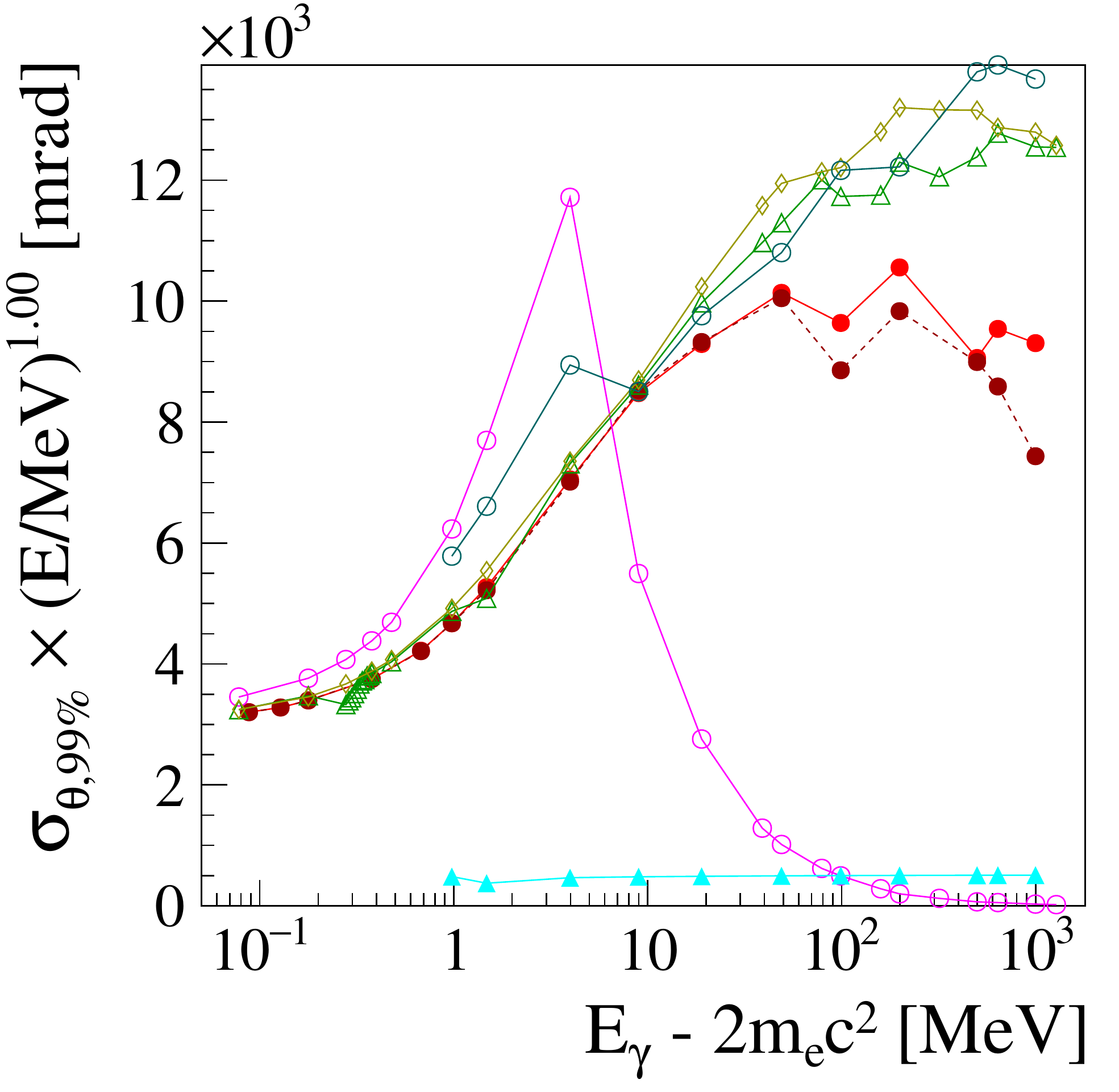}
  \caption{
    Angular resolution for the incident photon, assuming a perfect single track resolution.
    The $68\,\%$, $95\,\%$ and $99.7\,\%$ containment (resp. $1\sigma$, $2\sigma$ and $3\sigma$) can be roughly described at high energy by power laws of index $1.25$, $1.05$ and $1.00$ respectively for the exact model (\HELAS).
    \label{fig:compG4:res:model0}
  }
\end{center}
\end{figure}

Figure~\ref{fig:compG4:res:model0} shows the optimal (i.e. assuming a perfect tracking of the electron and positron) angular resolution $\sigma_{\theta}$ of single photons when the recoil nucleus cannot be measured.
For \HELAS\ the $68\,\%$ containment value can be fairly well described by a power law of index $1.25$, but it is modified by the form factor of Argon at high energies.
The $95\,\%$ and $99.7\,\%$ containment can be approximately described at high energy by power laws of index $1.05$ and~$1.00$ respectively.
\emstandard\ and \EGSa\ still give a poor description and largely underestimate the resolution.
The other models overestimate the resolution at high energy.

Multiple scattering is a dominating contributor to the angular resolution in pair telescopes.
If we consider, for example, a homogeneous tracker with optimal tracking, the single track resolution can be estimated by
\begin{equation}
\sigma_{\theta, \rm track} = (p/p_1)^{-3/4},
\end{equation}

where $p_1$ is a characteristic momentum that parametrises a particular detector
\cite{Bernard:2012uf,Bernard:2013jea}.

\begin{figure}[\plotflag]
\begin{center}
  \includegraphics[width=0.45\linewidth]{ResModel_Ar_cont68_select2_power1_0.pdf}
  \includegraphics[width=0.45\linewidth]{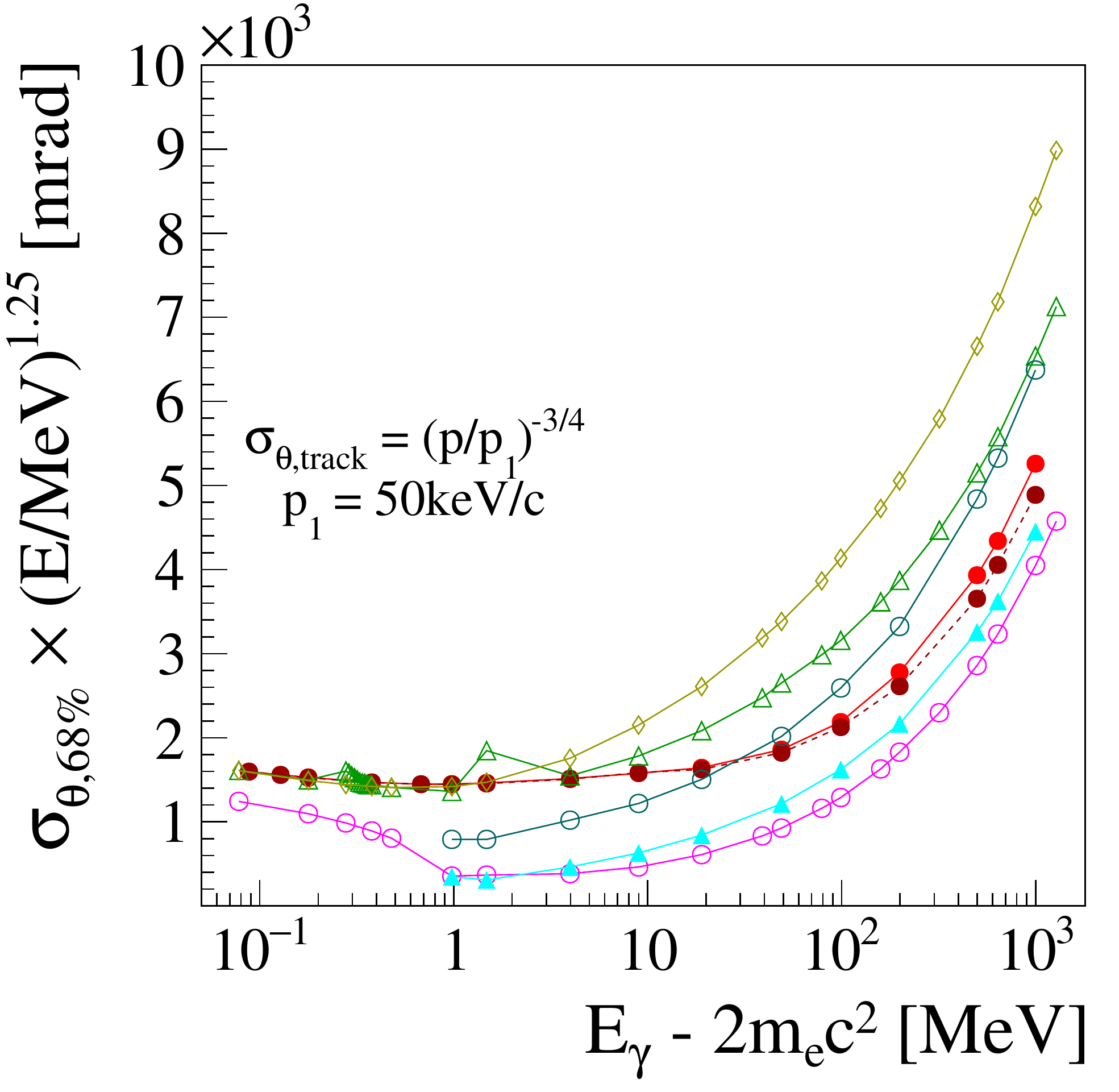}
  \includegraphics[width=0.45\linewidth]{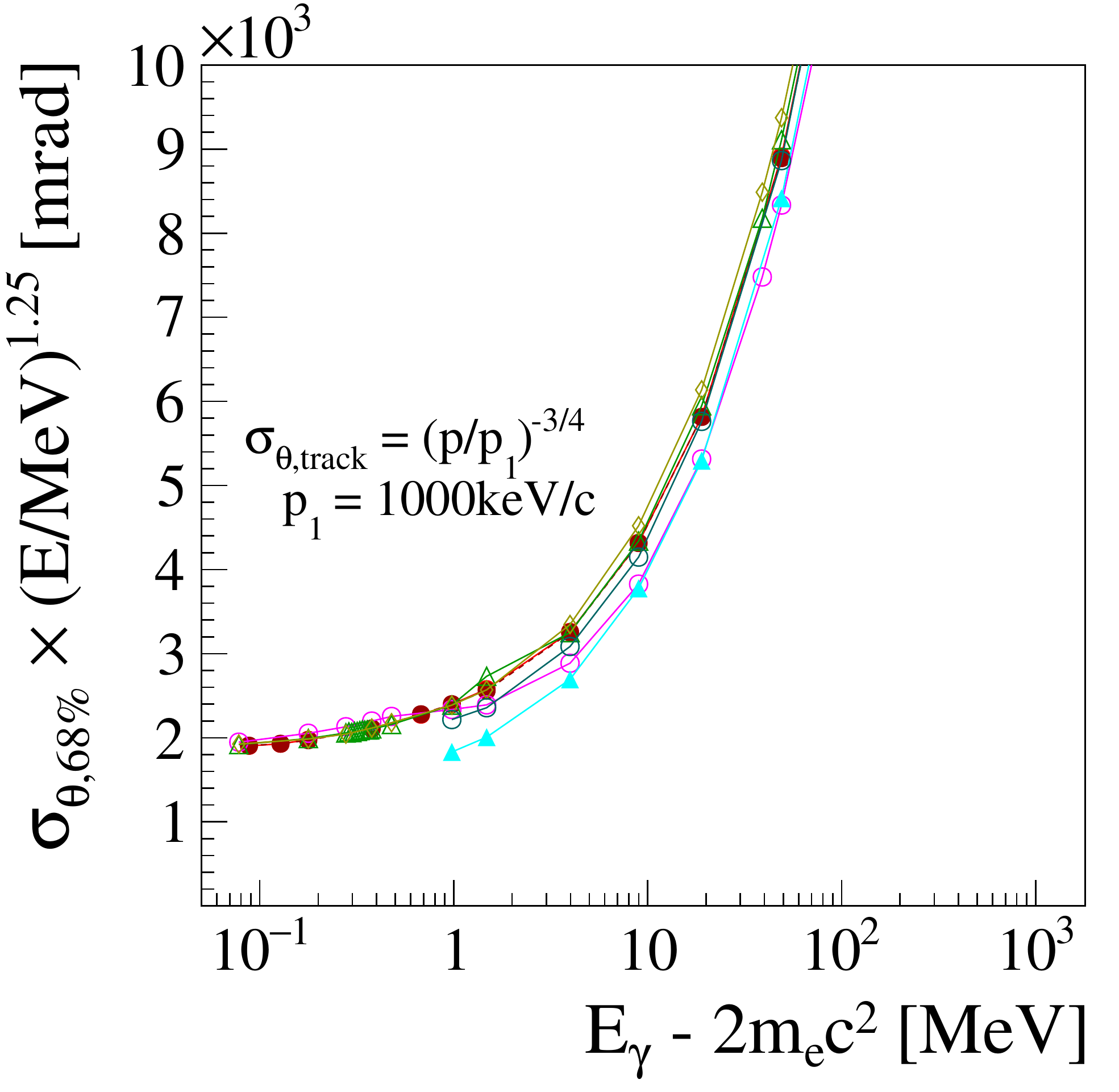}
  \caption{
    Angular resolution taking into account a single track resolution $\sigma_{\theta, \rm track} = (p/p_1)^{-3/4}$, with $p_1=50\,\kilo\electronvolt/c$ (middle) and $p_1=1\,\mega\electronvolt/c$ (right).
    The left plot recalls Fig.~\ref{fig:compG4:res:model0}, left, as a reference of the situation with perfect tracking.
    \label{fig:compG4:res:angular}
  }
\end{center}
\end{figure}

Figure~\ref{fig:compG4:res:angular} shows the resulting resolution for $p_1=1\,\mega\electronvolt/c$, which describes a typical tungsten-silicon telescope, and $p_1=50\,\kilo\electronvolt/c$, which approximately describes gaseous telescopes.
Past and present telescopes with tungsten converters, such as the \LAT, were not affected by the inaccuracy of existing generators.
On the other hand, future telescopes with minimised scattering will need to use accurate models. 

\begin{figure}[\plotflag]
\begin{center}
  \includegraphics[width=0.45\linewidth]{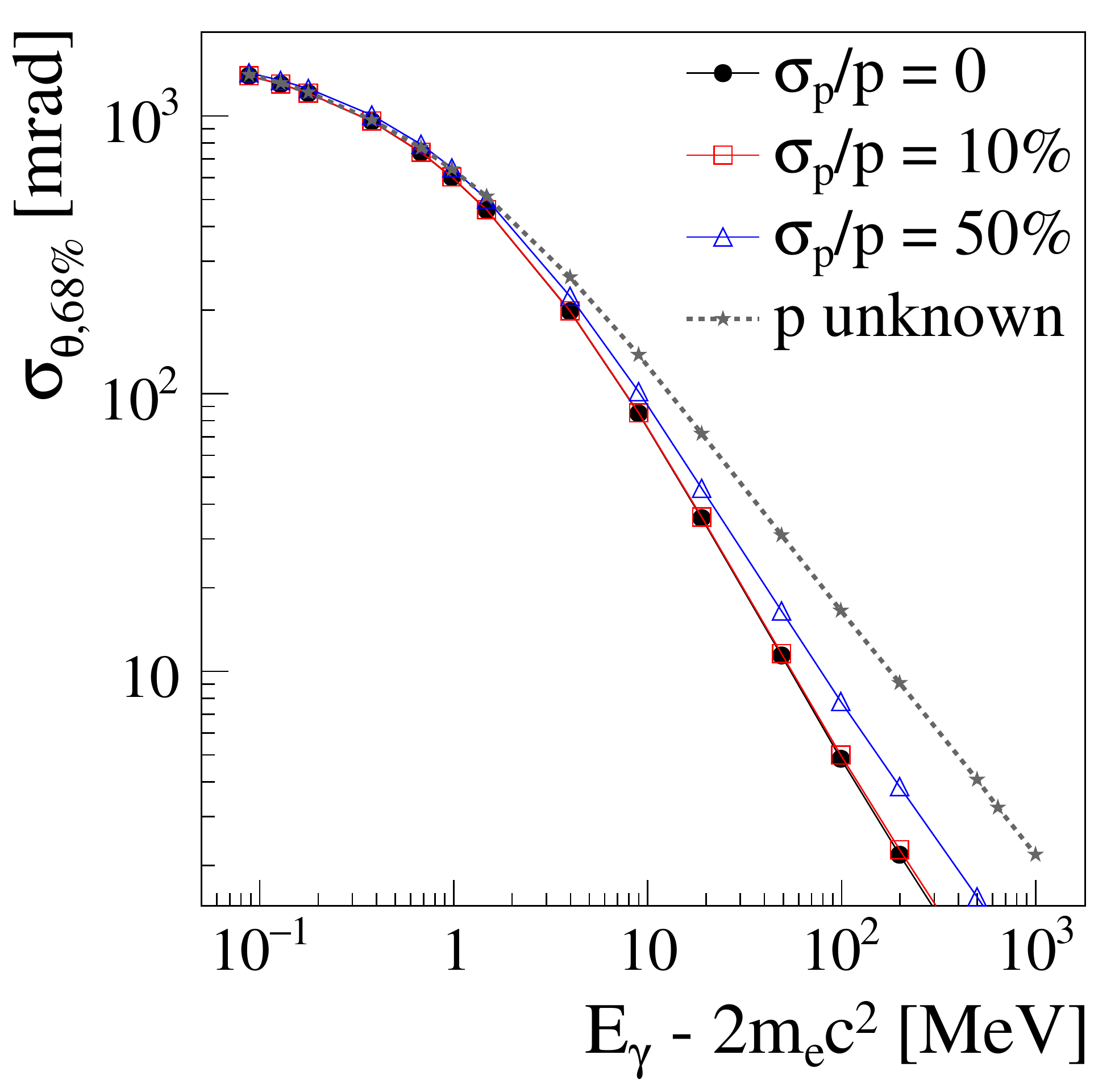}
  \caption{
    Angular resolution, using the \HELAS\ model, taking into account the energy resolution for the electron and positron $\sigma_{p}/p$.
    When $p_{i}$ is unknown, the photon direction is estimated by the bisector of the electron and positron directions.
    At high energy, the resolution becomes worse in absence of momentum information.
    An energy resolution of 50\,\% allows to recover most of the optimal resolution.
    \label{fig:compG4:res:energy}
  }
\end{center}
\end{figure}

The direction of the photon is estimated by the sum of the 4-momenta of the measured electron and positron.
This requires the knowledge of the energy of the two leptons.
The energy resolution of the tracker has therefore an influence on the angular resolution for the photon.
In the extreme case where the energy is completely unknown, the photon direction is estimated averaging the two lepton directions.
Figure~\ref{fig:compG4:res:energy} shows the angular resolution of the photon for several values of the energy resolution $\sigma_{p}/p$.
At higher energies, a lack of information on the magnitude of the track momenta affects the angular resolution, with an increase of a factor 3 at 100\,\mega\electronvolt\ compared to the case where the momenta are perfectly known.
The angular resolution in that case is described by a power law of index 0.95 instead of 1.25.
A measurement of the momentum with a resolution $\sigma_{p}/p$ as high as 50\,\% is enough to recover most of this loss.

\subsection{Polarisation asymmetry}
\label{subsec:pol}

Polarimetry in the pair-conversion regime is a goal for several projects (e.g. HARPO~\cite{Gros:SPIE:2016} with a gaseous detector, e-ASTROGAM~\cite{E-Astrogam:2016} with a silicon tracker and even the \LAT~\cite{Giomi:2016brf}), and few generators take the photon polarisation into account.
Besides our exact 5D generator, only \livermorepola\ tries to describe the angular asymmetry in the case of polarised photons.

\begin{figure}[\plotflag]
\begin{center}
  \includegraphics[width=0.45\linewidth]{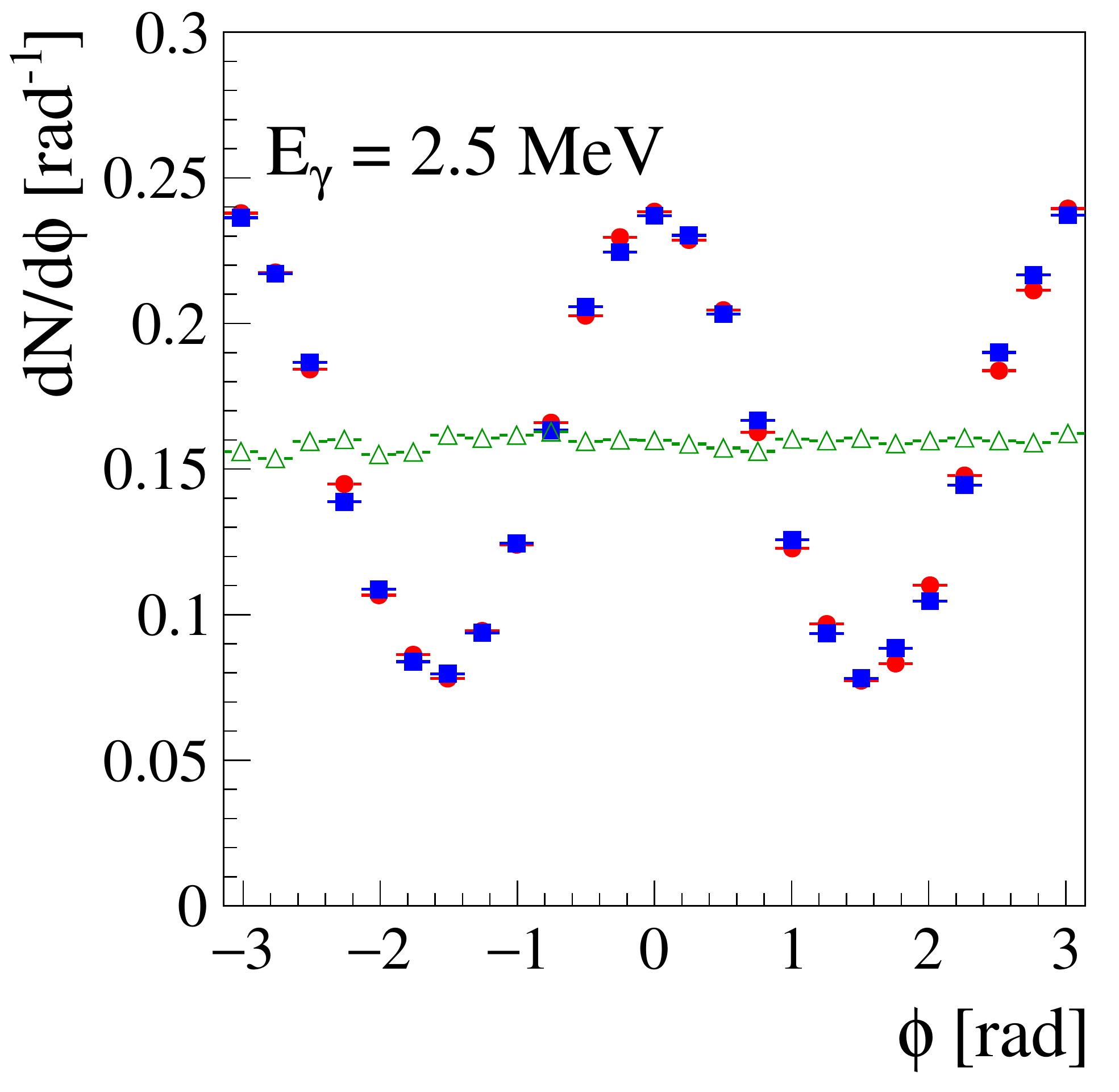}
  \includegraphics[width=0.45\linewidth]{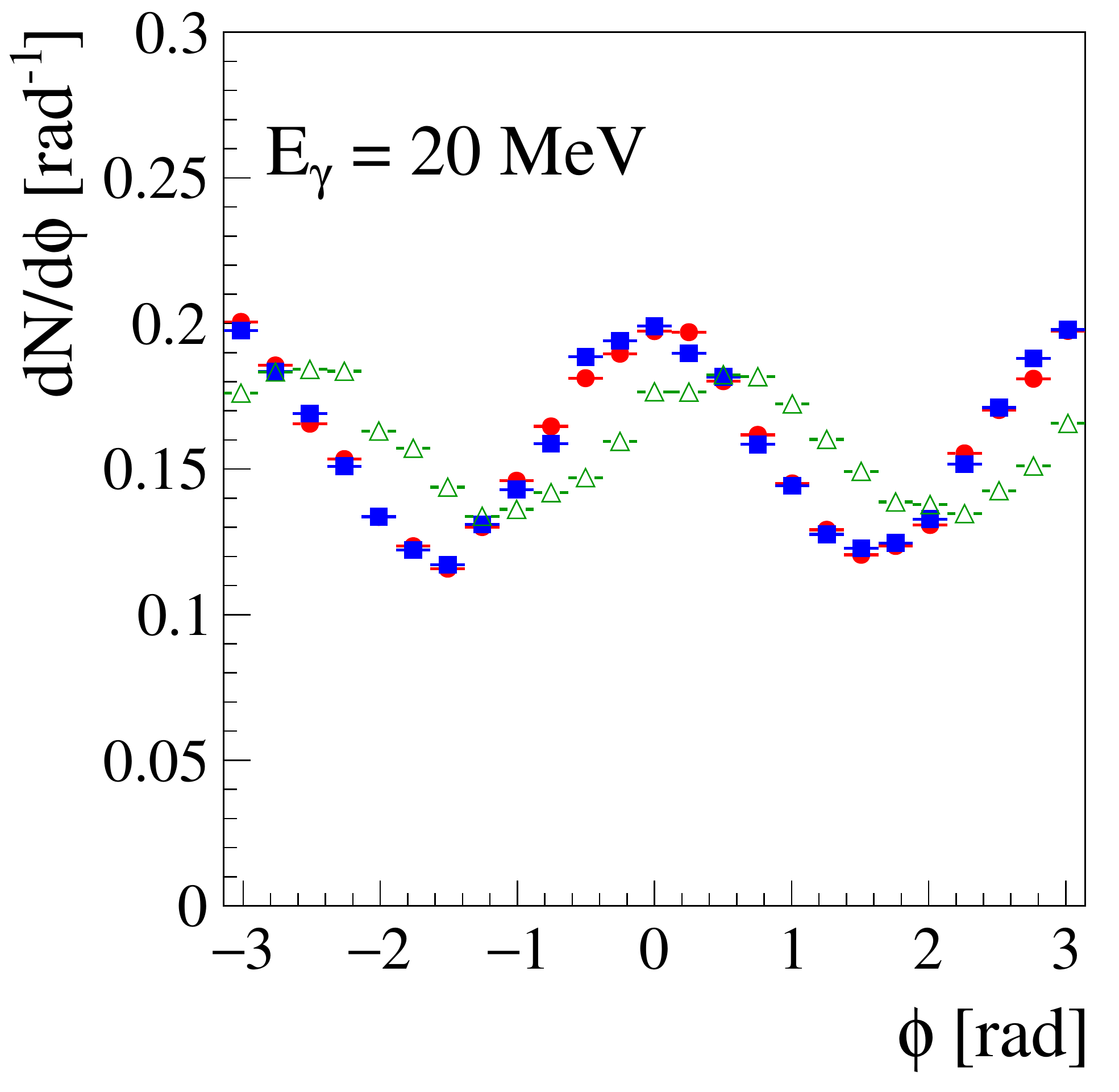}
  \includegraphics[width=0.45\linewidth]{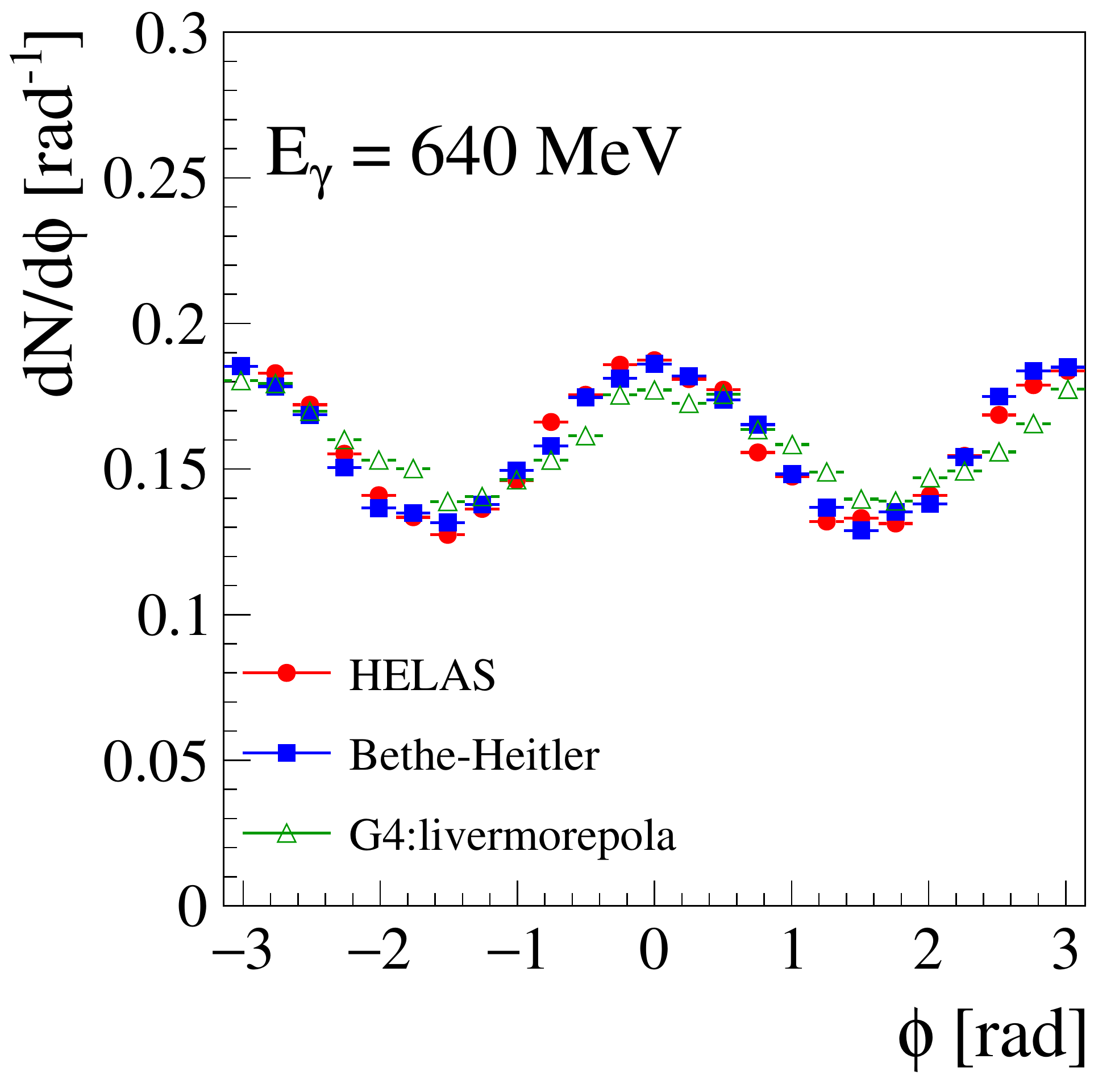}
  \caption{
    Distribution of the azimuthal angle $\phi$ for various energy of the incident photon, using the \HELAS\ (red circles), \BH\ (blue squares) and \livermorepola\ (green triangles) models.
    \HELAS\ and \BH\ show consistently a sine modulation, centred on $\phi_0 = 0$.
    \livermorepola\ shows a similar modulation, but with varying phase, and lower amplitude.
    The modulation disappears at low energy (2.5\,\mega\electronvolt), while it is maximum in the other models.
    \label{fig:compG4:omega:example}
  }
\end{center}
\end{figure}

Figure~\ref{fig:compG4:omega:example} shows examples of distributions of the azimuthal angle $\phi$ for three energy points.
We see that, as expected, \HELAS\ and \BH\ are still in good agreement.
Meanwhile, the amplitude and phase of the modulation in the case of \livermorepola\ is very different.

The azimuthal angle distribution for a fully linearly polarized beam is
\begin{equation}
  1+A\cos{\left(2(\phi-\phi_{0})\right)}
\end{equation}
We extract the parameters $A$ and $\phi_{0}$ by using the moments of the distribution~\cite{Gros:2016:azimuthal}:
\begin{eqnarray}
A&=&2\sqrt{ \langle \cos{2\phi}\rangle^2+ \langle \sin{2\phi}\rangle^2} ,
\\
\phi_0&=&\frac{1}{2}\arctan\left(\frac{ \langle \sin{2\phi}\rangle}{ \langle \cos{2\phi}\rangle}\right) .
\label{eq:fit}
\end{eqnarray}

\begin{figure}[\plotflag]
\begin{center}
  \includegraphics[width=0.45\linewidth]{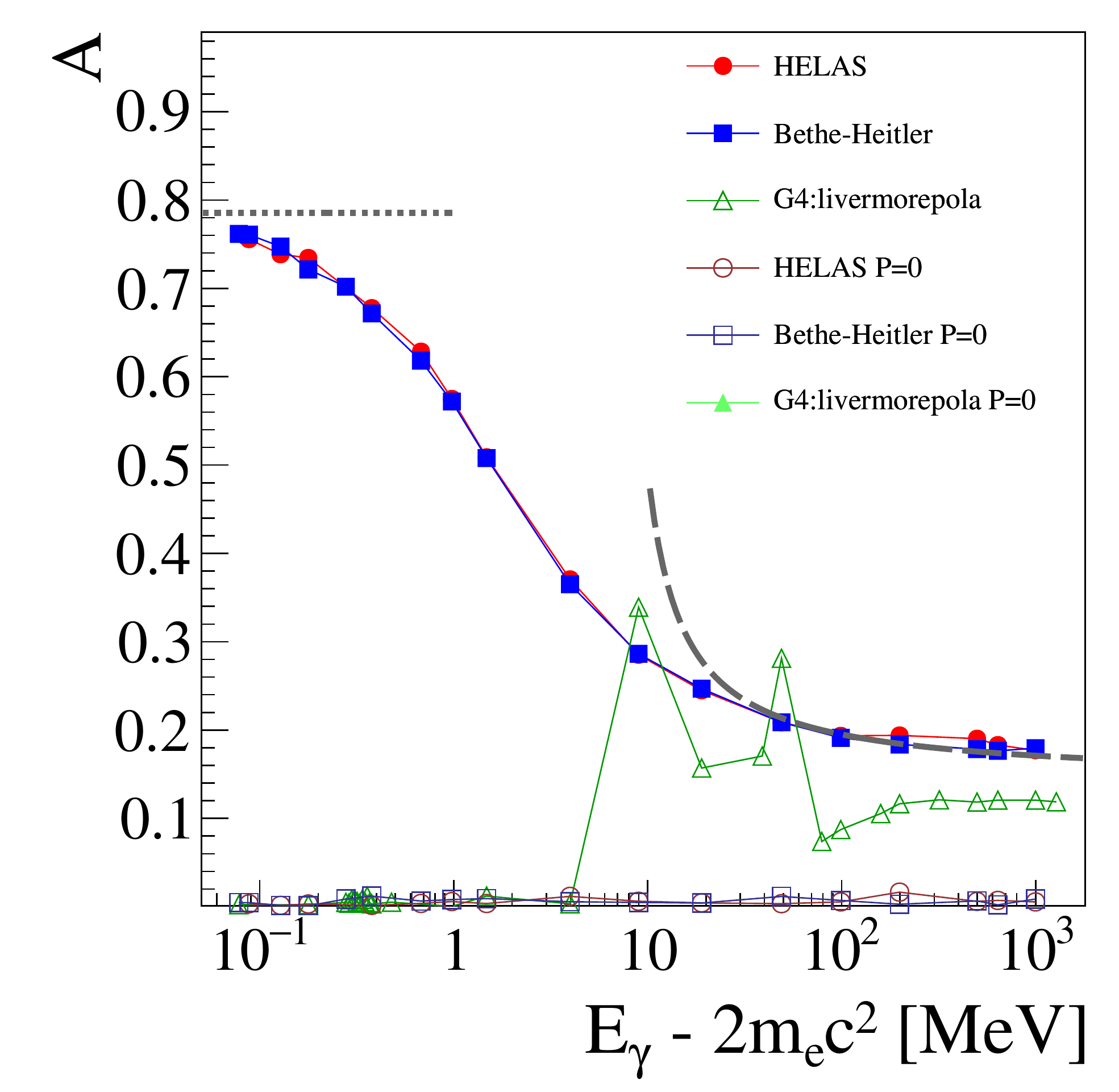}
  \includegraphics[width=0.45\linewidth]{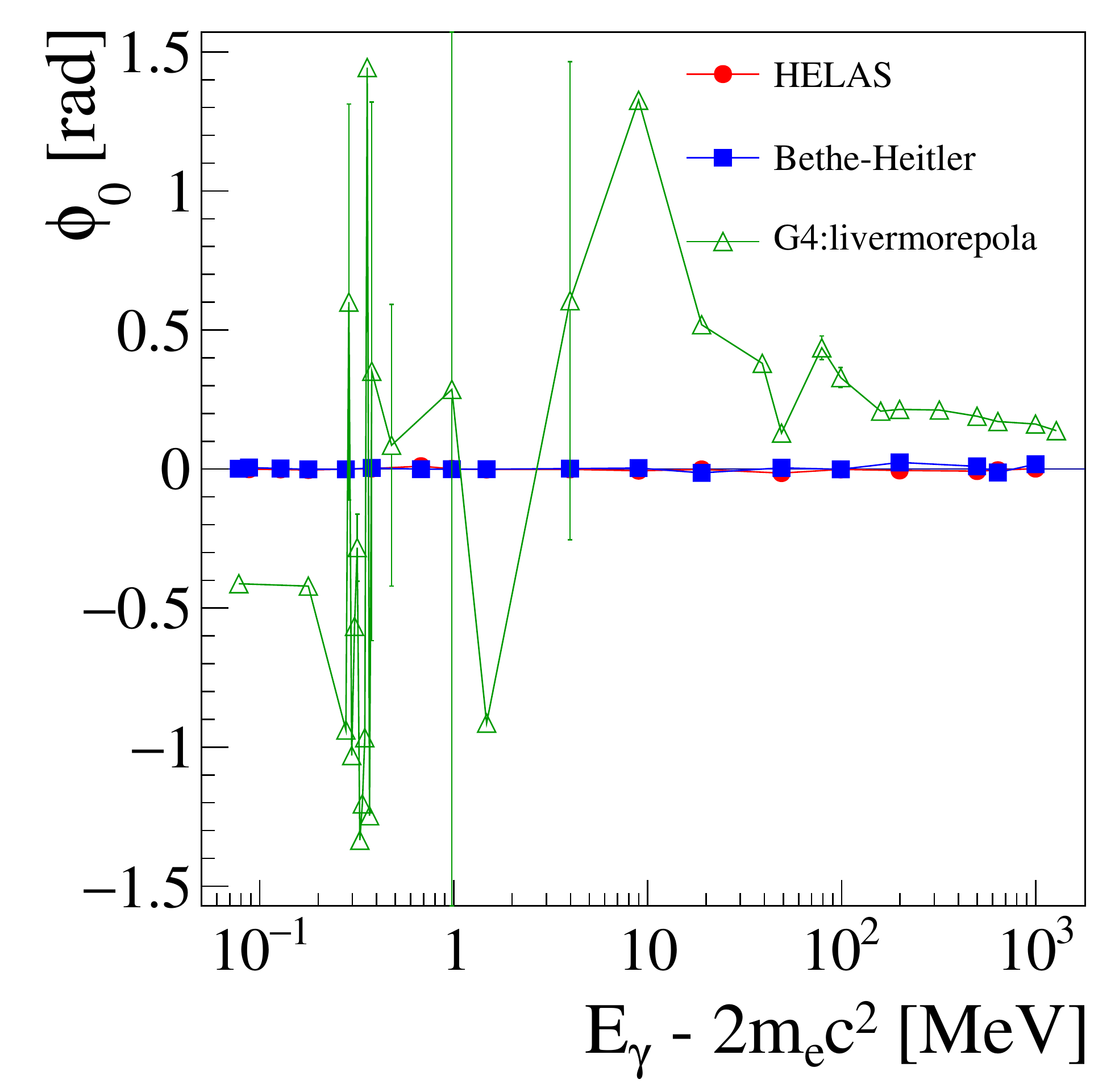}
  \caption{
    Amplitude $A$ and phase $\phi_0$ of the modulation of the distribution of the azimuthal angle $\phi$.
    When $A$ become too low ($E_{\gamma}<5\,\mega\electronvolt$ for \livermorepola), the phase $\phi_{0}$ loses meaning.
    The asymptotic values of $A$ at low and high energy are shown in dashed lines~\cite{Gros:2016:azimuthal}.
    The exact model agrees very well with these functions.
    The \livermorepola\ model fails to accurately describe the asymmetry, especially at low energy.
    \label{fig:compG4:Asym:model}
  }
\end{center}
\end{figure}

Figure~\ref{fig:compG4:Asym:model} shows the variations of $A$ and of $\phi_0$ with $E_{\gamma}$.
While all the results are consistent in the case of randomly polarised photons ($P=0$), the polarisation asymmetry for \livermorepola\ is clearly not correct under 100\,MeV.
Even at high energies, neither the amplitude, nor the direction are consistent with the exact generator.
At 100\,\mega\electronvolt, the exact generator gives an asymmetry $A=19.1 \pm 0.4$, while \livermorepola\ gives an asymmetry $A=8.7 \pm 0.6$, underestimating by a factor 2.

The form factor does not affect the polarisation asymmetry (plots not shown).

\section{Conclusion}

Pair conversion generators that are available in high-energy simulation frameworks rely on approximations that are sufficient for the description of electromagnetic showers in high-energy particle detectors, but they fail to give a correct description of the detailed kinematics of the pair production process, even of the 1D distributions of the main kinematic variables.
These inaccuracies are negligible for current detectors for which the multiple scattering dominates the angular resolution, but they become important for the new generation of telescopes.
We fully validated a 5D exact generator from 1\,\giga\electronvolt\ down to the pair production threshold.
We used it to obtain the values of the
1-$\sigma$,
2-$\sigma$ and
3-$\sigma$
angular-resolution containment values as a function of energy.

The photon polarisation is neglected in most available generators.
When it is taken into account, only a high energy approximation is used, while the polarisation asymmetry is actually maximal close to the conversion threshold.
The 5D exact generator is the only one that gives a polarisation asymmetry consistent with asymptotic values at low and high energies.

\section{Acknowledgements}

We acknowledge the support of the French National Research Agency (ANR-13-BS05-0002).


\begin{thebibliography}{99}

\bibitem{TIGRE:2001}
``The TIGRE gamma-ray telescope'',
T. J. O'Neill {\it et al.},
AIP Conf. Proc. 587, 882 (2001); 

\bibitem{MEGA:2005}
``Development and calibration of the tracking Compton/Pair telescope MEGA'',
G. Kanbach {\it et al.},
Nucl.\ Instrum.\ Meth.\ A {\bf 541} (2005) 310.

\bibitem{CAPSiTT:2010}
``CAPSiTT: Compton Large Area Silicon Timing Tracker for Cosmic Vision M3'',
F. Lebrun {\it et al.},
PoS(INTEGRAL 2010)034.

\bibitem{Morselli:2014fua} 
``GAMMA-LIGHT: High-Energy Astrophysics above 10 MeV,''
 A.~Morselli {\it et al.},
 Nuclear Physics B Proc. Supp. 239-240, 2013, 193,
 [arXiv:1406.1071 [astro-ph.IM]].

\bibitem{Moiseev:2015lva} 
 ``Compton-Pair Production Space Telescope (ComPair) for MeV Gamma-ray Astronomy,''
 A.~A.~Moiseev {\it et al.},
 arXiv:1508.07349 [astro-ph.IM].

\bibitem{PANGU:2015}
``PANGU: A High Resolution Gamma-Ray Space Telescope'',
 X. Wu {\it et al.},
PoS(ICRC2015)964.

\bibitem{E-Astrogam:2016}
``The e-ASTROGAM gamma-ray space mission'',
V. Tatischeff {\it et al.},
SPIE2016, 9905-91.

\bibitem{Caliandro:2013kba} 
 ``A new concept of y-ray telescope. LArGO: Liquid Argon Gamma-ray Observatory,''
 G.~A.~Caliandro {\it et al.},
 arXiv:1312.4503 [astro-ph.IM].

\bibitem{Aprile:2008ft} 
 ``Compton Imaging of MeV Gamma-Rays with the Liquid Xenon Gamma-Ray Imaging Telescope (LXeGRIT),''
 E.~Aprile {\it et al.},
 Nucl.\ Instrum.\ Meth.\ A {\bf 593}, 414 (2008)
 [arXiv:0805.0290 [physics.ins-det]].

\bibitem{Takahashi:2015jza} 
 ``GRAINE project: The first balloon-borne, emulsion gamma-ray telescope experiment,''
 S.~Takahashi {\it et al.},
 PTEP {\bf 2015}, no. 4, 043H01 (2015).

\bibitem{ber} 
``On the Detection of $\gamma$-Ray Polarization by Pair Production'',
T. H. Berlin and L. Madansky, 
Phys. Rev. {\bf 78}, 623 (1950).

\bibitem{Bernard:2013jea} 
 ``Polarimetry of cosmic gamma-ray sources above $e^+e^-$ pair creation threshold,''
 D.~Bernard,
 Nucl.\ Instrum.\ Meth.\ A {\bf 729}, 765 (2013)
 [arXiv:1307.3892 [astro-ph.IM]].

\bibitem{Gros:SPIE:2016}
``Measurement of polarisation asymmetry for gamma rays between 1.7 to 74 MeV with the HARPO TPC'',
P.~Gros {\it et al.},
SPIE2016, 9905-95.
[arXiv:1606.09417 [astro-ph.IM]].

\bibitem{Bernard:2012uf} 
 ``TPC in gamma-ray astronomy above pair-creation threshold,''
 D.~Bernard,
 Nucl.\ Instrum.\ Meth.\ A {\bf 701}, 225 (2013)
 Erratum: [Nucl.\ Instrum.\ Meth.\ A {\bf 713}, 76 (2013)]
 [arXiv:1211.1534 [astro-ph.IM]].

\bibitem{Gros:2016:azimuthal} 
 ``$\gamma$-ray polarimetry with conversions to e$^+$e$^-$ pairs: polarisation asymmetry and the way to measure it'',
 P.~Gros and D.~Bernard,
 accepted for publication by Astroparticle Physics,
 doi 10.1016/j.astropartphys.2016.12.006,
 arXiv:1611.05179 [astro-ph.IM].
 
\bibitem{Kawabata:1995th} 
``A New version of the multidimensional integration and event generation package BASES/SPRING,''
 S.~Kawabata,
 Comput.\ Phys.\ Commun.\ {\bf 88}, 309 (1995).

\bibitem{Lepage:1977sw}
``A New Algorithm for Adaptive Multidimensional Integration,''
  G.~P.~Lepage,
  J.\ Comput.\ Phys.\  {\bf 27} (1978) 192.

 
\bibitem{Heitler1954}
``The quantum theory of radiation'',
W. Heitler,
1954.

\bibitem{Murayama:1992gi} 
 ``HELAS: HELicity amplitude subroutines for Feynman diagram evaluations,''
 H.~Murayama, I.~Watanabe and K.~Hagiwara,
 KEK-91-11.

\bibitem{May1951}
``On the Polarization of High Energy Bremsstrahlung and of High Energy Pairs'',
 M. M. May,
Phys. Rev. 84, 265 - 270 (1951).

\bibitem{jau} 
{\it The theory of photons and electrons},
Jauch and Rohrlich, 
(Springer Verlag, 1976).

\bibitem{Jost:1950zz} 
 ``Distribution of Recoil Nucleus in Pair Production by Photons,''
 R.~Jost, J.~M.~Luttinger and M.~Slotnick,
 Phys.\ Rev.\ {\bf 80}, 189 (1950).

\bibitem{Geant4}
http://geant4.web.cern.ch/geant4/

\bibitem{Allison:2016lfl}
``Recent Developments in Geant4,''
 J.~Allison {\it et al.},
 Nucl.\ Instrum.\ Meth.\ A {\bf 835} (2016) 186.


\bibitem{Depaola:2000qd} 
 ``Azimuthal distribution for pair production by high-energy gamma-rays,''
 G.~O.~Depaola,
 Nucl.\ Instrum.\ Meth.\ A {\bf 452}, 298 (2000).

\bibitem{Baro}
``Analytical cross sections for Monte Carlo simulation of photon transport'',
J. Bar\'o {\it et al.},
Radiation Physics and Chemistry
 44 (1994) 531-552.

\bibitem{EGS5}
http://rcwww.kek.jp/research/egs/egs5.html

\bibitem{Hirayama:2005zm} 
 ``The EGS5 code system,''
 H.~Hirayama {\it et al.},
 SLAC-R-730, KEK-2005-8, KEK-REPORT-2005-8,
version: January 13, 2016.

\bibitem{Bielajew}
``Improved angular sampling for pair production in the EGS4 code system'',
 A. Bielajew,
PIRS-0287, 1991, revised version 1994.

\bibitem{Motz:1969ti} 
 ``Pair production by photons,''
 J.~W.~Motz, H.~A.~Olsen and H.~W.~Koch,
 Rev.\ Mod.\ Phys.\ {\bf 41}, 581 (1969).

\bibitem{Urban}
L.Urban in Geant3 writeup, section PHYS-211. Cern Program Library (1993).
%
%

\bibitem{Tsai:1973py} 
 ``Pair Production and Bremsstrahlung of Charged Leptons,''
 Y.~S.~Tsai,
 Rev.\ Mod.\ Phys.\ {\bf 46}, 815 (1974), Erratum: [Rev.\ Mod.\ Phys.\ {\bf 49}, 521 (1977)].

\bibitem{Giomi:2016brf}
 ``Estimate of the Fermi Large Area Telescope sensitivity to gamma-ray polarization,''
 M.~Giomi {\it et al.} [Fermi-LAT Collaboration],
 arXiv:1610.06729 [astro-ph.IM].

\end{thebibliography}
\end{document}